\documentclass[11pt]{article}
\usepackage[margin=2.5cm]{geometry}
\usepackage{amsmath}
\usepackage{amssymb}
\usepackage{bm}
\usepackage{graphicx}
\usepackage{mathtools}
\usepackage{booktabs}
\usepackage{xurl}
\usepackage{xspace}
\usepackage{textcomp}
\usepackage{microtype}
\usepackage{placeins}
\usepackage[numbers,sort&compress]{natbib}

\renewcommand{\eqref}[1]{Eq.~(\ref{#1})}

\newcommand{\ums}{\textmu{m/s}\xspace}
\newcommand{\um}{\textmu{m}\xspace}
\newcommand{\rads}{rad$^2\,\rm{s}^{-1}$\xspace}
\newcommand{\aff}[1]{\textsuperscript{#1}}

\title{Crowding limits relative bacterial surface accumulation with a density-induced effective heating}

\author{Yaocai Liu\aff{1,$\dagger$},
Rui He\aff{1,$\dagger$},
Da Wei\aff{2,$\dagger$,$*$},
Longfei Li\aff{2,$\dagger$},\\
Mingcheng Yang\aff{2,3,$*$},
Junhua Yuan\aff{1,$*$},
Yi Peng\aff{2,3,$*$}}

\date{}

\begin{document}

\maketitle

\begin{center}\small
\aff{1}Hefei National Laboratory for Physical Sciences at the Microscale and Department of Physics, University of Science and Technology of China, Hefei, Anhui, China\par
\aff{2}Beijing National Laboratory for Condensed Matter Physics, Institute of Physics, Chinese Academy of Sciences, Beijing 100190, China\par
\aff{3}School of Physical Sciences, University of Chinese Academy of Sciences, 19A Yuquan Road, Beijing 100049, China\par
\aff{$\dagger$}Y.L., R.H., D.W., and L.L. contributed equally to this work.\par
\aff{$*$}To whom correspondence should be addressed.\par
E-mail: weida@iphy.ac.cn, mcyang@iphy.ac.cn, jhyuan@ustc.edu.cn, pengy@iphy.ac.cn
\end{center}

\begin{abstract}
The accumulation of microswimmers near walls has been studied for decades. While hydrodynamics and single-cell kinematics are well-explored, the role of cell-cell interactions remains understudied. However, in realistic, crowded environments such as the human gut, these interactions can be just as crucial. Here, we demonstrate that they actively weaken bacterial surface accumulation. Analysis of entrapped cell tracks reveals that crowding enhances orientational fluctuations, which increase the effective rotational diffusion and facilitate wall escape. A minimal model shows that this collective effect is equivalent to raising the system's effective temperature with density.
\end{abstract}


\noindent\textbf{Significance:} Bacterial surface accumulation initiates biofouling, compromising submerged materials and medical implants. While decades of research have focused on single-cell hydrodynamics and behavior, the role of cell-cell interactions---crucial in realistic, crowded environments such as the human gut---remains less understood. Here, we show that crowding inherently weakens relative surface accumulation. This collective effect is quantified experimentally and captured by a minimal model as an increased effective rotational noise, or effective heating. This effective noise serves as a simple, powerful descriptor for the many-body interactions governing the accumulation in dense active systems, bridging single-cell mechanics and population-level ecology. The insights apply readily to other microswimmer systems, including microbot-assisted drug delivery, where controlling surface accumulation helps achieve sufficient local dosage.

\section{Introduction}
Materials submerged in aqueous environments become targets for microbial colonization within minutes~\cite{Vedie2021Review,Nurioglu2015biofouling}. Such microbial colonization is known as biofouling and results in undesirable consequences ranging from increased drag on ship hulls to infections on medical implants~\cite{Vedie2021Review}.
Bacteria, as primary colonizers, quickly adhere to surfaces and initiate biofilm formation. This initial surface accumulation is the critical first step that creates a scaffold for secondary colonization by larger organisms~\cite{Qian2022biofilmFurtherFouling}.

Decades of research has sought to resolve the mechanisms of bacterial surface accumulation~\cite{Frymier1995,Berke2008,Li2009,Bianchi2017,Junot2022}. Current research confirms that the accumulation of bacteria at interfaces arises from a competition between boundary-attracting and boundary-escaping mechanisms. Key factors promoting accumulation include the alignment effect during collisions between rod-shaped bacteria and walls~\cite{Drescher2011,Li2009}, hydrodynamic attraction arising from the force dipole component of the cell's flow field~\cite{Berke2008,Schaar2015}, and biological surface appendages like fimbriae that enhance adhesion~\cite{Perez2019,Kimkes2020BacteriaHairReview}. Conversely, active reorientation mechanisms such as bacterial tumbling~\cite{Bianchi2017}, passive angular noise~\cite{Berg1993}, and hydrodynamic effects of force quadrupole~\cite{Wei2025} facilitate escape from wall entrapment.

Despite this progress, a critical factor remains underexplored: the role of short-range cell-cell interactions---both steric and hydrodynamic---in wall escape. This gap stems from a research focus on elucidating individual bacterial behavior and near-surface fluid dynamics, which requires tracking cells at very low concentrations ($n \ll 0.1n_0$, where $n_0 = 8 \times 10^8$ cells/ml). As a benchmark, bacterial concentrations in the human gut are $0.1$--$1~n_0$~\cite{Macfarlane1998,Breton2016}. While neglecting cell-cell interactions was justified in the dilute limit considered in earlier studies, understanding accumulation in a more realistic context (i.e., with abundant other microswimmers around) demands their inclusion. These close-range interactions may not only introduce reorientations that aid wall escape but also trigger complex behaviors like L\'evy walks~\cite{Ariel2017}.

In this work, we investigate bacterial surface accumulation in chambers (thickness $H \approx 100$~\um) across two decades of concentrations ($10^{-1}-10^{1}~n_0$). Interestingly, we find that surface accumulation quickly diminishes (i.e., the bulk becomes nearly equally populated) once the concentration exceeds $\sim1~n_0$. To explore this phenomenon, we perform experiments with different bacterial strains, which vary in body length and tumbling behavior. Nevertheless, the density-induced weakening of relative surface accumulation is consistently observed. This density-induced effect is captured by an increasing effective temperature for cell-body rotation in a minimal model. These results strongly suggest that bacterial collisions, rather than individual motility traits or hydrodynamics, dominate boundary accumulation at concentrations exceeding $1~n_0$. Our findings highlight that, in densely populated natural and industrial settings such as the gut and bioreactors, cell-cell interactions need to be considered as one of the primary factors in microbe-surface dynamics.

\section{Results}

\subsection{Surface accumulation diminishes in dense suspensions}
\begin{figure}
 \centering
 \includegraphics[width=1.0\textwidth]{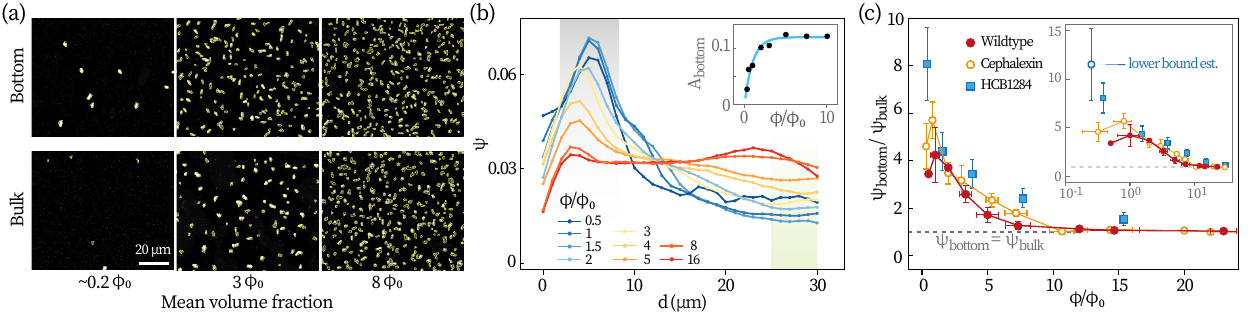}
 \caption{\textbf{Surface accumulation in bacterial suspensions of different densities.} (a) Confocal microscopy images near the bottom (upper row) and near the center (lower row) at different bacterial concentrations $\phi$. (b) The normalized area fraction $\psi$ as a function of height $d$ from the chamber bottom at varying $\phi$. Inset: The 2D area fraction near the bottom, $A_{\rm bottom}$, versus $\phi$. (c) Accumulation level characterized by $\psi_{\rm bottom}/\psi_{\rm bulk}$ for varying $\phi$ and bacterial strains, with the inset presenting the same data on a semi-log scale. Markers and error bars represent the mean and SD from $N \geq 3$ experiments. The empty marker in the inset indicates an estimated lower bound (measurements with effectively zero $\psi_{\rm bulk}$ excluded).}
 \label{fig:accumulation}
 \vspace{-2mm}
\end{figure}

We observe bacterial suspensions in fluidic chambers using confocal microscopy. Videos at varying heights $d$ from the bottom are acquired. Images are contrast-adjusted and binarized, and cells are detected via a thresholding algorithm~\cite{Tinevez2017Trackmate} (see \textit{Methods}). Typical processed images are presented in Fig.~\ref{fig:accumulation}(a). The upper and lower rows show images taken near the bottom wall ($d\approx3$~\um) and in the bulk ($d\approx30$~\um), respectively. We observe that accumulation near the top and the bottom surfaces is qualitatively the same~\cite{Wei2025} but bottom images are clearer (especially under higher concentrations). We therefore focus on the region $d \in [0, d_0]$ near the bottom wall, with $d_0 = 30$~\um.

We examined suspensions of different bacterial strains at volume fractions $\phi$ ranging from $0.1$ to $20\,\phi_0$ (0.019\%--3.9\%), where $\phi_0 = 0.19\%$ is the unit volume fraction corresponding to the unit number density $n_0$ for typical \textit{E. coli} cells. Before each experiment, $\phi$ is determined directly from optical density measurements. Because the number density $n$ is more sensitive to cell size than $\phi$, and the strains we use differ considerably in length, we adopt $\phi$ to characterize concentration hereafter (see \textit{Methods}).

We begin with wildtype \textit{E. coli} cells. In dilute suspensions ($\phi \lesssim 3\phi_0$), bacteria accumulate near the no-slip walls, in agreement with previous reports~\cite{Berke2008,Li2009,Molaei2014}. At very low densities (e.g., $\phi\approx0.2\phi_0$), cells are heavily concentrated near the bottom, leaving the bulk nearly empty of cells [Fig.~\ref{fig:accumulation}(a)]. At an intermediate concentration (e.g., $\phi = 3\phi_0$), both the bulk and bottom become more populated, though the bottom remains denser; the visual difference between the two regions is less pronounced. In crowded suspensions (e.g., $\phi = 8\phi_0$), the bulk and bottom densities become indistinguishable.
The vertical population distribution is characterized by $A(d)$, the absolute 2D area fraction of cell bodies at height $d$, and its normalized profile $\psi(d) \equiv A(d)/\int_0^{d_0} A(z){\rm d}z$. Typical profiles of $\psi$ are shown in Fig.~\ref{fig:accumulation}(b). The cell concentration peaks several \um above the bottom [gray shaded area in Fig.~\ref{fig:accumulation}(b)]. This is consistent with recent observations and is understood through a combination of the cells' hydrodynamic and steric interactions with the surface~\cite{Vladescu2014droplet,Burriel2024,Htet2024hovering,Wei2025}. For clarity, values referenced for the \textit{bottom} correspond to the maximum value within this bottom region, while values referenced for the \textit{bulk} correspond to the mean for $d\in[25,30]$~\um [green shaded in Fig.~\ref{fig:accumulation}(b)].

In the dilute regime ($\phi \lesssim 3\phi_0$), the $\psi(d)$ profile remains qualitatively similar with increasing concentration $\phi$ [blue traces in Fig.~\ref{fig:accumulation}(b)]. Despite this stable normalized profile, the absolute area fraction in the bottom region, $A_{\rm bottom}$, increases and saturates exponentially [Fig.~\ref{fig:accumulation}(b) inset]. We further characterize surface accumulation by the ratio $\psi_{\rm bottom}/\psi_{\rm bulk}$ from the profile $\psi(d)$ [see Fig.~\ref{fig:accumulation}(c)], where a larger ratio indicates stronger accumulation. Consistently, this ratio is stable against varying $\phi$ in the dilute regime and diminishes in dense suspensions, see Fig.~\ref{fig:accumulation}(c) and inset.

Note that the ratio $\psi_{\rm bottom}/\psi_{\rm bulk}$ quantifies surface accumulation in a relative sense; its decrease in dense suspensions reflects that the bulk becomes increasingly populated as the surface saturates, consistent with the observations in Fig.~\ref{fig:accumulation}(a). This trend follows a clear physical logic. When the surface becomes crowded, enhanced cell-cell interactions could significantly facilitate cell reorientation, increasing the detachment rate.
Consequently, surface accumulation weakens, yielding a more homogeneous distribution.

We hypothesize that the stable trend of $\psi_{\rm bottom}/\psi_{\rm bulk}$ in the dilute regime arises because density-induced detachment is negligible compared to the baseline rate from single-cell tumbling, a dominant mechanism in extremely dilute suspensions~\cite{Frymier1995,Molaei2014,Junot2022}.

To test this hypothesis, we examine surface accumulation in tumble-free \textit{E.~coli} (HCB1284; see \textit{Methods}), shown as blue data in Fig.~\ref{fig:accumulation}(c). These cells display a monotonic decrease of $\psi_{\rm bottom}/\psi_{\rm bulk}$ with $\phi$: accumulation is strongest in the most dilute suspensions. For $\phi\leq0.2\phi_0$, the bulk is often empty within our observation span, resulting in an effectively infinite $\psi_{\rm bottom}/\psi_{\rm bulk}$. These data are excluded from the reported statistics since the ratio is undefined when the bulk contains no cell; the corresponding lower bound is shown as an open marker in Fig.~\ref{fig:accumulation}(c) inset. The results support our hypothesis: when the baseline detachment rate provided by single cell tumbling is absent, increasing density consistently boosts detachment and hence weakens surface accumulation.

The tumble-free strain is longer [$3.5 \pm 1.2$~\um, mean $\pm$ standard deviation (SD)] than the wildtype ($2.8 \pm 0.6$~\um)\footnote{Although the mean body length of HCB1284 with mCherry inserts (3.5~\textmu{m}) is only 25\% longer than the wildtype (2.8~\textmu{m}), the former has larger variability [1.2~\textmu{m} standard deviation (SD)] than the latter (0.6~\textmu{m}), resulting in that the smooth swimming bacterial suspensions contain much larger fractions of long cells than wildtype suspensions contain.}. To ensure this size difference does not cause the different $\psi_{\rm bottom}/\psi_{\rm bulk}$ trend, we study drug-elongated wildtype cells. A mild dose of cephalexin is applied to bacteria (see \textit{Methods}) and yields cells of lengths $6.0 \pm 1.3$~\um. We confirm that their swimming speed and tumble behaviors are unaffected---though their angular persistence after tumbling is stronger. The surface accumulation of these elongated cells [orange data in Fig.~\ref{fig:accumulation}(c)] remains qualitatively the same as wildtype.

\subsection{Angular kinematics of entrapped bacterial motion}

\begin{figure}[htbp]
 \includegraphics[width=0.48\textwidth]{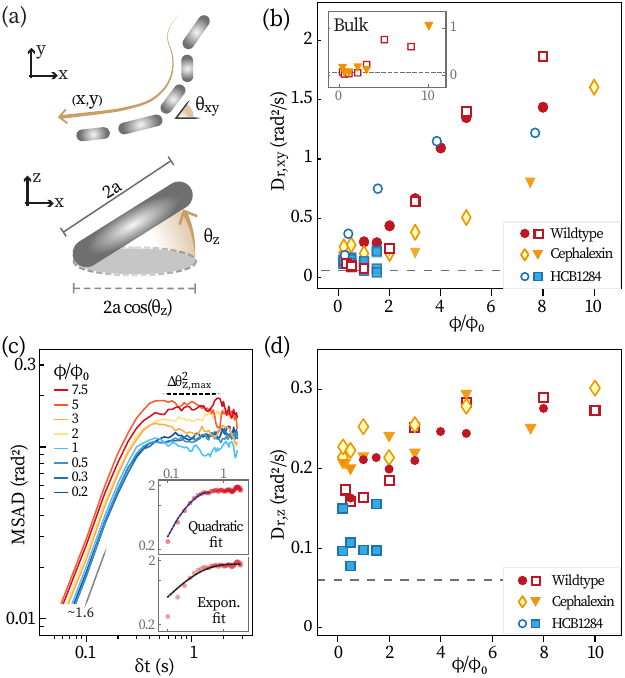}
 \caption{\textbf{Angular noise of bacteria near a surface.} (a) Schematic of cell tracking in the $xy$-plane (top) and how polar angle $\theta_z$ can be inferred from apparent length (bottom). (b) In-plane rotational diffusivity, $D_{r,xy}$, as a function of $\phi$. Different symbols denote different experiments; inset shows bulk measurements. (c) Mean square angular displacement (MSAD) of $\theta_z$ across concentrations. Insets illustrate the two methods for extracting the polar rotational diffusivity $D_{r,z}$ (see \textit{Methods}). (d) $D_{r,z}$ as a function of $\phi$. Dashed lines in (b) and (d) indicate the thermal background of $0.06~\rm{rad^2/s}$.}
 \label{fig:Dr}
\end{figure}

We next study the dynamics of bacteria swimming near surfaces to obtain a microscopic perspective on how cell-cell interactions affect surface accumulation. Since a cell's entrapment at and escape from the wall are dictated by its swimming angle, we focus on the angular kinematics of entrapped cells. A cell is considered entrapped if it continuously remains within $d\leq 5$~\um of the bottom (see \textit{Methods}).

We first characterize the in-plane angular noise by the effective rotational diffusivity $D_{r,xy}$. Entrapped cells are tracked, and from the track direction $\theta_{xy}(t)$ [Fig.~\ref{fig:Dr}(a)] we compute the mean square angular displacement (MSAD). For a purely diffusive process, MSAD scales linearly with the time interval $\delta t$ ($\sim 2 D_{r,xy} \delta t$). However, due to hydrodynamic wall effects, entrapped cells swim in circular paths near the surface~\cite{Lauga2006}, which introduces a convective contribution $d\theta_{xy} \sim \omega_{\rm circ} \delta t$ and hence a quadratic term in the MSAD. Here $\omega_{\rm circ}$ is the constant angular speed of circular swimming. To isolate the contribution of diffusion, we fit the MSAD quadratically to $\omega_{\rm circ}^2 \delta t^2 + 2D_{r,xy} \delta t + \mathrm{const.}$ and extract $D_{r,xy}$ from the coefficient of the linear term; see \textit{Methods}. This method consistently captures experimental results; a comparison with alternative methods is provided in the \textit{Supporting Information} (SI).

Figure~\ref{fig:Dr}(b) displays results of different strains. For all strains, $D_{r,xy}$ increases with $\phi$, showing an approximately linear dependence. Bulk measurements yield quantitatively similar results [Fig.~\ref{fig:Dr}(b) inset]. In the dilute limit, all strains approach the background thermal noise, estimated as $k_B T/8\pi\eta a^3=0.06~\rm{rad^2/s}$ for a passive particle~\cite{Berg1993}, where $k_B$ is the Boltzmann constant, $T$ is room temperature, $\eta$ is the viscosity of water, and $a=1.5$~\um is a typical bacterial half-length.

Compared to the in-plane angular kinematics, kinematics of the polar angle [$\theta_z$ in Fig.~\ref{fig:Dr}(a), lower panel] is more directly involved in wall entrapment and detachment~\cite{Htet2024hovering,Bianchi2017,Wei2025}. However, tracking the evolution of $\theta_z$ is challenging, especially in crowded environments. Here, we employ an approximate yet effective technique to address this issue. When tracking the cells' motion in the $xy$-plane, we also record their 2D-projected length in real-time as $L(t)$, which is related to $\theta_z$ as $L(t)=2a\cos{\theta_z}$ [Fig.~\ref{fig:Dr}(a), lower panel]. For each individual cell, its length $2a$ is approximated as the maximum value of $L(t)$, denoted as $L_{\rm max}$. Note that, as $L_{\rm max} \leq 2a$, results obtained by this method tend to overestimate $\theta_z$; and longer cell bodies lessen this relative error.

We track $\theta_z$ from eligible entrapped tracks (see \textit{Methods}) and compute the corresponding MSAD. Representative data for elongated wildtype cells at varying densities are displayed in Fig.~\ref{fig:Dr}(c). At short time intervals ($\delta t \leq 0.3$~s), MSAD exhibits superdiffusive scaling with time across all $\phi$. At longer time scales, the MSAD plateaus at a value $\Delta \theta^2_{z,\mathrm{max}}$ that increases with $\phi$, where $\Delta \theta_{z,\mathrm{max}}$ corresponds to the maximum polar angle variation allowed for entrapped cells.

The superdiffusivity primarily arises from the cell body wobbling. When a bacterium swims, its cell body counter-rotates against its flagellar bundle, causing the polar angle $\theta_z$ to vary with time approximately sinusoidally: $d \theta_z \sim \sin{(\omega_{\rm wob} \delta t)}$, where $\omega_{\rm wob}$ is the wobbling frequency [see detailed experimental measurements in~\cite{Bianchi2017,Kamdar2023wobble}]. At short times, this yields a convective dependence ($d\theta_z\sim \omega_{\rm wob} \delta t$) in addition to the existing angular diffusion in $\theta_z$, culminating in the observed superdiffusive behavior of the MSAD. Note that this coexistence of convective and diffusive angular variations is analogous to the case of $D_{r,xy}$, justifying the use of the same protocol to extract the rotational diffusion coefficient $D_{r,z}$. We thus perform quadratic polynomial fitting up to the plateau to obtain $D_{r,z}$ [Fig.~\ref{fig:Dr}(c), upper inset]. Results across strains are displayed in Fig.~\ref{fig:Dr}(d).

The information given by wall entrapment, i.e., the plateaus at longer times of MSAD, provides a complementary view and corroborates the measured $D_{r,z}$. To characterize this feature, which quadratic fitting cannot capture, we simply approximate the wall entrapment as a harmonic potential in $\theta_z$ (see \textit{Methods}). This approximation predicts MSAD to exhibit an exponential saturation---from which $D_{r,z}$ can also be extracted [Fig.~\ref{fig:Dr}(c), lower inset]. Results from both methods agree well, differing by only $\sim$10\%. We report $D_{r,z}$ from the quadratic polynomial fit for methodological consistency with $D_{r,xy}$ (see \textit{SI} for details). Like $D_{r,xy}$, $D_{r,z}$ increases with $\phi$ but with a slower trend~[Fig.~\ref{fig:Dr}(d)]. At high concentrations, $D_{r,z}$ amounts to $\sim$20\% of $D_{r,xy}$. Note that because cells with the strongest fluctuations tend to escape and are not sampled, our results represent a lower bound for the actual $D_{r,z}$ near the surface. These results directly show how surface crowdedness intensifies polar angle fluctuations.

\begin{figure}[h]
\includegraphics[width=0.48\textwidth]{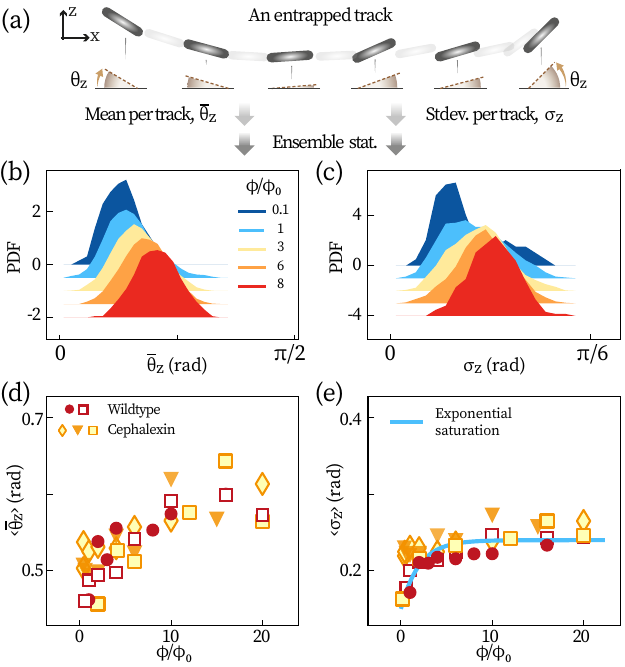}
 \caption{\textbf{Polar angle fluctuation of bacteria near a surface.} (a) Definition of metrics: $\overline{\theta}_z$ and $\sigma_z$ denote the temporal mean and standard deviation of $\theta_z$ for an individual track. Probability distributions of $\overline{\theta}_z$ (b) and $\sigma_z$ (c) for different concentrations; data are vertically offset for clarity. (d,e) Ensemble averages $\langle\overline{\theta}_z\rangle$ and $\langle\sigma_z\rangle$ corresponding to the distributions in (b) and (c), respectively. Symbols represent different experiments. The blue line in (e) is an exponential saturation with the same characteristic rate as in Fig.~\ref{fig:accumulation}(b) inset.}
 \label{fig:thetaz}
\end{figure}

We further substantiate the polar angle kinematics with more straightforward statistics of $\theta_z$. For $\theta_z(t)$ of each entrapped track, we compute its mean $\overline{\theta}z$ and standard deviation $\sigma_z$ over time~[Fig.~\ref{fig:thetaz}(a)] and examine their distributions across concentrations. Representative data for elongated cells are displayed in Figs.~\ref{fig:thetaz}(b) and (c), with each distribution comprising $O(10^2\text{--}10^3)$ tracks. Consistent with the increase of $D_{r,z}$ with $\phi$ [Fig.~\ref{fig:Dr}(d)], entrapped cells experience stronger fluctuations in $\theta_z$ at higher concentrations, as seen in the right-shifting distributions of $\sigma_z$ in Fig.~\ref{fig:thetaz}(c). Meanwhile, we find that entrapped cells also swim with a more tilted orientation (larger $\overline{\theta}_z$) at higher concentrations [Fig.~\ref{fig:thetaz}(b)]. The increasing trends of $\langle\overline{\theta}_z\rangle$ and $\langle\sigma_z\rangle$ with $\phi$ are shown in Figs.~\ref{fig:thetaz}(d) and (e), respectively, where $\langle \cdot \rangle$ denotes the ensemble average. Data for non-elongated wildtype cells (red symbols) exhibit qualitatively the same trend. Interestingly, the saturation of $\langle\sigma_z\rangle$ with $\phi$ appears to follow a similar exponential form [blue line in Fig.~\ref{fig:thetaz}(e)] as that of $A_{\rm bottom}$ [Fig.~\ref{fig:accumulation}(b) inset]. In other words, the crowdedness of the surface appears to enhance its dwellers' polar angle fluctuation proportionally.

\begin{figure}[htb!]
\includegraphics[width=0.96\textwidth]{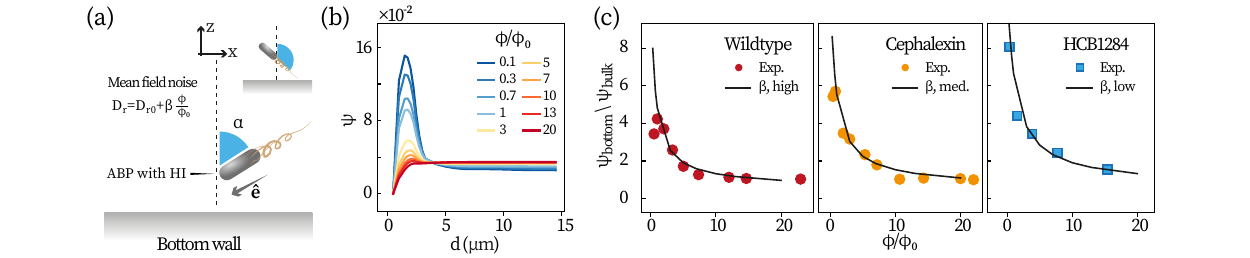}
 \vspace{-2mm}
 \caption{\textbf{Modeling results}. (a) Schematic of the model, defining the adjusted polar angle $\alpha=\pi/2-\theta_z$ for cells approaching the surface and $\alpha=\pi/2+\theta_z$ for departing cells (inset). (b) Normalized distribution $\psi(d)$ at different concentrations. (c) Surface accumulation ratio $\psi_{\rm bottom}/\psi_{\rm bulk}$ from simulations with different noise-density coupling strengths $\beta$. Symbols are experimental data; the baseline rotational diffusivity is $D_{r0} = 0.4$~\rads for all simulations; $\beta=$0.51, 0.38, and 0.26 for panels in (c) from left to right, respectively.}
 \label{fig:model}
\end{figure}

\subsection{Modeling results}
A minimal mean-field model successfully captures how bacterial density affects surface accumulation. Density-induced effects are encompassed by an increasing effective temperature for cell-body rotation.
Bacteria are modeled as active Brownian particles (ABPs) in the $xz$-plane, confined between no-slip walls at $z=0$ and $H$ [Fig.~\ref{fig:model}(a)], with periodic boundary conditions along the $x$-axis (length $W$). They are subject to a rotational white noise whose intensity increases with cell density. The model includes far-field hydrodynamic interactions (HI) and steric exclusion between the cell and the wall. Each particle is described by its position $\bm{r} = (x, z)$ and orientation $\hat{\bm{e}} = (-\sin{\alpha}, -\cos{\alpha})$ (for left-moving particles). The angle $\alpha$ represents an adjusted polar angle: $\alpha=\pi/2 - \theta_z$ for cells approaching the surface [Fig.~\ref{fig:model}(a)] and $\alpha=\pi/2 + \theta_z$ for cells leaving [Fig.~\ref{fig:model}(a) inset]. The ABP kinematics are governed by:
\begin{equation}\label{eq:model}
\begin{aligned}
\dot{\bm{r}} &= v_0 \hat{\bm{e}} + u_z(\alpha, z) \hat{\bm{z}} - \tfrac{1}{\gamma_T} \nabla_{\bm{r}} U;\\
\dot{\alpha} &= - \Omega(\alpha, z) + \sqrt{2D_{r,z}} \xi.
\end{aligned}
\end{equation}
Here, $v_0$ is the constant self-propulsion speed. Note that the particle's actual speed also includes contributions from steric and hydrodynamic interactions, and thus differs from $v_0$. $u_z$ is the HI-induced velocity along the $z$-axis (unit vector $\hat{\bm{z}}$), $\gamma_T$ is the translational drag coefficient, and $U$ is a short-range potential for steric wall exclusion. $\Omega$ is the HI-induced angular velocity, and $\xi$ is a Gaussian white noise with $\langle \xi(t) \rangle=0$ and $\langle \xi(t)\xi(t') \rangle=\delta(t-t')$, where $\delta(\cdot)$ is the Dirac delta function.

Hydrodynamic interactions between the particle and the wall are approximated by modeling the particle as a force dipole $p$ following Ref.~\cite{Berke2008}:
\begin{equation}
\begin{aligned}
u_z\! &= -\tfrac{3p}{64\pi \eta z^2}(1 \!-\! 3\cos^2{\alpha}), \\
\Omega &= -\tfrac{3p \cos{\alpha} \sin{\alpha}}{64\pi \eta z^3} \left[ 1 + \tfrac{1}{2}\tfrac{\epsilon^2 \! - \! 1}{\epsilon^2 \! + \! 1}(1 \!+\! \cos^2{\alpha}) \right],
\end{aligned}
\end{equation}where $\epsilon$ is the particle aspect ratio. These expressions describe interactions near the bottom wall; for the top wall, $z$ is replaced by $H-z$ and the $z$-component of $\hat{\bm{e}}$ is reversed.

The particle further interacts with the wall through steric exclusion and wall alignment. Steric exclusion is implemented via a short-range harmonic potential $U(d) = K (d-\ell)^2/2$ for $d < \ell$ (zero otherwise), where $\ell$ is the characteristic length for steric interactions and approximates the bacterial body width. Wall alignment is modeled by resetting the polar angle $\alpha$ to $\pi/2$ when $d < \ell$ and $\alpha < \pi/2$. For more numerical details see \textit{Methods}.

Finally, density-induced effects are incorporated via a mean-field rotational noise $D_{r}(\phi) = D_{r0} + \beta \phi/\phi_0$, where the slope $\beta$ quantifies the noise increase per unit bacterial density. The linear dependence of $D_{r}$ on $\phi$ is a first-order approximation based on our experimental measurements near the surface and in the bulk [Fig.~\ref{fig:Dr}(b)], yielding $\beta \sim 0.1$~\rads. This approximation is also supported by particle-based simulations of self-propelled rods (SPRs). The simulations show that far-field hydrodynamic interactions (dominant in the dilute regime) and steric collisions (dominant in the dense regime) each yield a (near-)linear dependence (Fig.~S5 in the \textit{SI Appendix}). Further simulation details of the SPR model are provided in the \textit{Methods}.

The weakening of surface accumulation with density is captured by our model. With typical parameters---$a=2$~\um, $v_0=20$~\ums, $p=0.6$~pN$\cdot$\um~\cite{Drescher2011}, $H=100$~\um, $D_{r0}=0.4$~\rads, and $\beta=0.51$~\rads---the model reproduces the evolution of $\psi(d)$ with $\phi$: from a peak near the surface to a nearly flat distribution [Fig.~\ref{fig:model}(b)]. Furthermore, by adjusting only the noise-density coupling strength $\beta$, the model reproduces the distinct trends for all strains [Fig.~\ref{fig:model}(c)]. The results reasonably show that $\beta$ is highest for the normal-sized wildtype and lowest for the tumble-free strain. The model also provides reasonable predictions for the residence time $\tau_{\rm res}$ of the tumble-free strain at low concentrations. The best fit to measurements is $D_{r0}=0.1$~\rads and $\beta=0.15$~\rads. At $\phi=0.25$, 0.4, and 1.5~$\phi_0$, the modeled (measured) values are: $\tau_{\rm res}=21.1$~s (23.7~s), 13.4~s (7.4~s), and 3.1~s (3.1~s), respectively.

In the minimal model, the increase of $D_r$ with density is imposed rather than derived. To examine the robustness of the minimal model
, and to explore its microscopic basis, we consider three classes of extended models (SI Appendix). First, free-space simulations model bacteria as hydrodynamically interacting force dipoles, as sterically interacting SPRs (each rod is a rigid two-sphere dimer with anisotropic friction), or as rods subject to both hydrodynamic and steric interactions; these test whether the interactions alone, without wall confinement, produce the linear increase of $D_r$ with density (Fig.~S5).
Second, a mean-field SPR model, with wall repulsion and alignment generated by dimer-wall contact while $D_r(\phi)$ remains assigned, tests the role of finite particle geometry (Fig.~S6).
Third, near-wall multiparticle simulations explicitly include pairwise hydrodynamic interactions and wall-image flows, with or without a short-range cell-cell scattering torque (Fig.~S7). In these simulations, the intrinsic single-particle diffusivity $D_{r0}$ is held fixed, and the density-dependent effective rotational noise emerges from interactions rather than being prescribed (Fig.~S7). These models show consistent results. In free space, hydrodynamic interactions, steric collisions, and their combination each produce a near-linear increase of the effective rotational diffusivity with density (Fig.~S5), supporting the first-order closure $D_r=D_{r0}+\beta\phi/\phi_0$. The SPR model preserves the qualitative decay of relative surface accumulation, showing that the result is insensitive to finite rod geometry and the detailed implementation of wall alignment (Fig.~S6). More directly, the near-wall multiparticle simulations reproduce the decay of $\psi_{\mathrm{bottom}}/\psi_{\mathrm{bulk}}$ without imposing a density-dependent $D_r$ (Fig.~S7). Finally, the decay trends for all three experimental strains and all model classes are qualitatively the same (Fig.~S8). These results indicate that different microscopic mechanisms can be coarse-grained into a common effective heating in the rotational degree of freedom, while the microscopic details mainly determine how strongly density enhances that noise. The minimal model captures this mechanism in its simplest form, which also underpins its general applicability.

\section{Discussion}
Our study reveals that bacterial surface accumulation weakens significantly as concentrations increase beyond the dilute regime. This transition marks a shift from a state governed by single-cell tumbling to one dominated by density-induced collective interactions. Crucially, the net effect of these complex interactions can be semi-quantitatively captured by a single parameter---an enhanced effective rotational noise in a minimal mean-field model.

This density-dependent mechanism is often overlooked in studies focusing solely on single-cell hydrodynamics, yet it is critical for predicting accumulation under realistic, crowded conditions such as in the human gut. Quantitatively, our results establish that peer interaction becomes the governing mechanism of bacterial surface accumulation for $\phi \gtrsim 1.0~\phi_0$ [Fig.~\ref{fig:accumulation}(c) and inset]. In this regime, our data support the following picture: as density increases, the surface population saturates [Fig.~\ref{fig:accumulation}(b) inset] while the bulk gradually fills, homogenizing the overall distribution and reducing relative surface accumulation ($\psi_{\rm bottom}/\psi_{\rm bulk}$). The characteristic scale for this exponential saturation is also approximately $1.0~\phi_0$, as evidenced by both population and noise level data [Figs.~\ref{fig:accumulation}(b) inset and \ref{fig:thetaz}(e)].

Some remarks are necessary regarding the highly simplified model we employ. Several detailed known mechanisms are deliberately omitted, including near-field hydrodynamics beyond the dipole approximation~\cite{Zhang2021PNAS,Leishangthem2024,Wei2024}, cell body wobbling~\cite{Kamdar2023wobble}, and flagellar hook flexibility~\cite{Tao2025PNAS}. Nevertheless, the model proves effective. This effectiveness underscores a key insight: in dense suspensions, the effects of the neglected mechanisms, combined with frequent cell-cell interactions (e.g., collisions), reduce to the effect of a simple white noise. This reduction is analogous to a central limit theorem for reorientation kinematics, making a mean-field enhancement of rotational noise a powerful and sufficient descriptor for population-level behavior. Although the modeling results are quantitatively reasonable, we emphasize that the model's primary aim is to provide qualitative insight into surface accumulation in dense suspensions.

This framework applies broadly to other microswimmers, such as algae, sperm, and artificial microbots, where particle geometry and flow fields can be more complex and single-particle behavior more elusive. Our results suggest that such complexity can be dissolved in many-body interactions, advancing understanding of surface accumulation in general. The common decay trend of the accumulation ratio across strains and models (Fig.~S8) further suggests that the effective heating in the rotational degree of freedom is a generic consequence of crowding, largely independent of the microscopic details of the interactions. This has important implications for microswimmer-based applications. For example, in microbot-assisted drug delivery, surface accumulation of drug-loaded particles promotes wall-assisted transport~\cite{wang2025reviewRollingWall} and enables sufficient local dosage~\cite{liu2022DrugDeliveryLocalDose}; understanding and modulating this accumulation is therefore of key interest.

\section{Materials and Methods}

\subsection{Bacteria culture}
\textit{E. coli} strain HCB1284 and BW25113 are used for this study.
Smooth swimming strain HCB1284 ($\Delta CheY$) is transformed with pBAD33 \textit{flic} and pTrc99a mCherry or only transformed with pBAD33 \textit{flic}. Bacteria is cultured in TB medium with appropriate antibiotics (25~\textmu{g}/ml chloramphenicol, 100~\textmu{g}/ml ampicillin) at 33~$^\circ \rm{C}$ to an optical density (OD, 600~nm wavelength) of 0.51. Arabinose with final concentration of $0.01\% (w/w)$ is added to express extra flagella at the beginning of the culture and IPTG with final concentration of 100 \textmu{M} were added at OD of 0.23 to express mCherry.

The wildtype strain BW25113 expresses green fluorescent protein. This strain is first cultured overnight at 37.0~$^\circ \rm{C}$ in TB medium, shaken at 200 rpm. The saturated culture is diluted 100~fold in fresh TB medium and incubated at 30~$^\circ \rm{C}$ under identical shaking conditions. After 6.5~h, bacteria are harvested by centrifugation (1000 or 800$\times$g, 5~min), washed, and resuspended in motility buffer (MB) (10~mM potassium phosphate, 67~mM NaCl, and 0.1~mM EDTA). To induce cell elongation, cephalexin (Solarbio Inc., $\sim$30~\textmu{g}/ml) is added, 3.5~h after the start of the fresh culture. The resulting cells exhibit body lengths of $6.0 \pm 1.3$~\um [mean $\pm$ SD].

\subsection{Volume fill fraction}
Concentration of bacterial suspensions is determined by measuring the optical density at 600~nm (OD$_{600}$), which primarily reflects the volume fraction of cell bodies. For wild-type \textit{E.~coli} (considered as a cylinder of length $2a = 3$~\um and diameter $2b = 1$~\um), an OD$_{600} = 1$ corresponds to a number density $n = 8 \times 10^8$~cells/ml $\equiv n_0$ and a volume fraction $\phi = n \cdot 2\pi a b^2 = 0.19\% \equiv \phi_0$.
Because bacterial strains of varying lengths are used, and their number density varies at the same OD$_{600}$, we use $\phi$ to characterize concentration. In experiments, after centrifugation, dense suspensions reach $\phi = 40$--$80\,\phi_0$, which are then diluted to working concentrations of $0.1$--$20\,\phi_0$; beyond this range, bacterial turbulence emerges~\cite{Wei2024}.

\subsection{Observation}
Suspensions of the strain BW25113 of varying concentrations are filled into flow chambers and sealed with UV-cured glue (NOA81). Chambers are custom made ($H\approx85$~\um) with a slide and a coverslip spaced by double-sided tape. Confocal microscopy is performed at varying heights $d$ (resolution 1~\textmu{m}) from the bottom walls with a Prime-95B camera at a frame rate of 20~fps. A 40$\times$ objectives is used (0.275~\um/pixel).

The observation chambers for the smooth swimming strain HCB1284 are fabricated by attaching a PDMS sheet to a coverslip with double-sided tapes. The resultant chamber depth is 100~\textmu{m}. The same strain with or without mCherry are mixed in different ratios, allowing us to inspect single-cell dynamics even in dense suspensions. Before observation, samples are placed on the stage for 5~minutes to reach steady state. Nikon confocal microscopy equipped with a 40$\times$ water immersion objective (NA 1.15, 0.275~\um/pixel) is used to obtain the image every 1~\textmu{m} from the bottom of the chamber. For measuring the residence time of the bacteria staying at surface, we used Nikon Ti-E microscopy equipped with a 561 nm laser and a 40x objective (NA 0.6); the videos were recorded at a frame rate of 25~fps with a scientific CMOS camera (Photometric kinetix).

\subsection{Image processing}
We apply a band-pass filter (3--50 px) to the confocal images to get rid of the background pixel intensity variation as well as the salt and pepper noise. The image is then adjusted by histogram saturation and binarized before further analysis such as cell detection.

After binarization, the number of occupied pixel $N_{\rm occu}$ reports the two dimensional (2D) fill ratio (by bacterial cell bodies) within the imaging plane $A(d)$, as $A(d)\equiv\langle N_{\rm occu} \rangle/N_{\rm field}$. Here $\langle\cdot\rangle$ represents average over $\mathcal{O}(10^2)$ frames of images. $N_{\rm occu}$ is computed as the total area enclosed by detected cell body contours. Contours shorter than 9~px (corresponding to the typical cross-sectional size of a bacterium) are considered noise and excluded. $N_{\rm field}$ is the total number of the pixels per image. From $A(d)$, the scaled probability density function (PDF) $\psi(d) \equiv A(d)/\int_0^{d_0} A(z){\rm d}z$ is computed.

Cell detection and tracking are performed with the TrackMate plugin in ImageJ~\cite{Tinevez2017Trackmate}. Further analyses (e.g., curvature, diffusion, etc.) are carried out with custom Python programs.

\subsection{Definitions of $\theta_{xy}$ and $\theta_z$}
Videos used for obtaining single-cell kinematics near the bottom surface are recorded at 20--25~fps. Recordings last 5~s or 30~s. $\theta_{xy}$ is defined based on the a cell's 2D track. An eligible track for the analysis of MSAD must (1) last longer than 1.0~s---which ensures meaningful temporal analysis; and (2) have a mean speed of more than 5~\ums---which excludes stuck and dead cells. Using the cell's body orientation $\theta^{\rm body}_{xy}$ changes the rotational diffusivity $D_{r,xy}$ obtained negligibly (see \textit{SI}). We therefore have opted for using $\theta_{xy}$.

Meanwhile, the polar angle $\theta_z$ is defined as the included angle between the long axis of the cell body and the bottom surface. A tracks is eligible for statistical analysis if it meets all three of the following criteria: (1) lasts longer than 0.5~s; (2) has a mean speed more than 5~\ums; and (3) has at least an aspect ratio of 3---which guarantees the accuracy of detecting the cell's body orientation.

\subsection{Extracting $D_{r,xy}$}
Consider a surface-entrapped bacterium swimming circularly at an angular speed $\omega_{\rm circ}$ and subjected to an in-plane rotational diffusivity $D_{r,xy}$. Over a time interval $\delta t$, its measured directional change is $d\theta_{xy} = \omega_{\rm circ} \delta t+\delta\theta_{xy}+\zeta$, where $\delta\theta_{xy}$ is the diffusive angular change and satisfies $\langle \delta\theta_{xy}^2 \rangle = 2D_{r,xy} \delta t$, and $\zeta$ denotes measurement uncertainty. Assuming the three terms are uncorrelated, the MSAD reads $\langle d\theta_{xy}^2 \rangle= \omega_{\rm circ}^2 \delta t^2+\langle\delta\theta_{xy}^2\rangle+\langle \zeta^2 \rangle = \omega_{\rm circ}^2 \delta t^2+2D_{r,xy}\delta t+\mathrm{Const.}$, where $\langle \cdot \rangle$ denotes ensemble average. $D_{r,xy}$ is taken as half the linear coefficient from a least-square quadratic polynomial fitting. For more details of fitting methods, see \textit{SI}.

\subsection{Extracting $D_{r,z}$}
The quadratic polynomial fitting approach is also applied to obtain $D_{r,z}$. In this case, the MSAD in the small-to-intermediate time scales (i.e., before MSAD plateaus) is fitted to $ \langle d\theta_z^2 \rangle= \omega_{\rm circ}^2 \delta t^2+2D_{r,z}\delta t+\mathrm{Const.}$, where $\omega_{\rm wob}$ denotes the angular speed of cell body wobbling projected onto the $xz$-plane.

Meanwhile, to exploit information on the long time scales (the plateaued regime), we develop a model that helps extracting $D_{r,z}$ by exponential fitting. In this model, the wall-entrapment is approximated as a harmonic potential that exerts a recovery torque on $\theta_z$:
\begin{equation}\label{eq:MSADexpon}
\gamma \dot{\theta}_z = -k (\theta_z - \Theta_z) + \xi_z.
\end{equation}
Here $k$ denotes the stiffness of the harmonic potential, $\gamma$ is the rotational drag coefficient of a bacterium within $xz$-plane, $\Theta_z$ is the equilibrium position, and $\xi_z$ is a white noise. $\xi_z$ satisfies $\langle \xi_z(t)\xi_z(t') \rangle=2 D_{r,z} \gamma^2 \delta(t-t')$ with $\delta(\cdot)$ the Dirac delta function.
By definition, $\mathrm{MSAD}\equiv \langle [\theta_z - \theta_z(0)]^2 \rangle$, we first integrate Eq.~(\ref{eq:MSADexpon}) from $t'=0$ to $t'=t$ to obtain an analytic expression for $\theta_z(t)-\theta_z(0)$; and then perform ensemble average on the square of the terms. With $\langle \xi_z \rangle=0$, $\langle \theta_z(0) \rangle=\Theta_z$, and the expression of $\langle \xi_z(t)\xi_z(t') \rangle$, we obtain the expression describing the saturating profile:
\begin{equation}
 \mathrm{MSAD} = D_{r,z}\,\tau (1 - e^{\frac{-2\delta t}{\tau}}),
\end{equation}
with $\tau=\gamma/k$ the characteristic time scale of MSAD saturation. $\tau$ and $D_{r,z}$ are the only two fitting parameters in this approach. One may further include cell body wobbling into Eq.~(\ref{eq:MSADexpon}) but we find that the fitting results change negligibly.

\subsection{Numerical details of the minimal modeling}
Simulation parameters are set based on experimental measurements. The simulation box dimensions are $W \times H = 100 \times 100$~\um, and the fluid viscosity is $\eta = 0.89$~mPa$\cdot$s. the self-propulsion speed is $v_0 = 20$~\ums, the force dipole strength is $p = 0.6$~pN$\cdot$\um~\cite{Drescher2011}, and the aspect ratio is $\epsilon = 5$. For simplicity, anisotropy in the translational friction coefficient $\gamma_T$ induced by the cell body shape is neglected; $\gamma_T$ is approximated as that of a sphere of effective radius $r = 2$~\um, i.e., $\gamma_T = 1/6\pi\eta r$. For computational efficiency, hydrodynamic interactions are computed only for $d < 0.15H$ from each wall (i.e., a total region of $0.3H$ across the channel height); varying this cutoff from $0.06H$ to $0.3H$ affects the numerical results negligibly ($<10\%$). For computing steric interactions, the wall steric potential stiffness is $K = 10$~pN/\um, and the characteristic length for steric interactions is $\ell = 1.2$~\um.

\subsection{Simulation details of SPR suspensions}
Particle-based SPR simulations are performed to rationalize the linear dependence of the effective rotational diffusivity $D_{r}$ on $\phi$ used in the minimal mean-field model. SPR suspensions are simulated under periodic boundary conditions. Each SPR (representing a bacterium) is modeled as a hard-sphere dimer with a self-propulsion velocity $v_0$, with a force dipole of strength $p$ embedded along its swimming orientation ${\hat {\bm e}}$. The equations of motion for particle $i$ are
\begin{align}
 {\dot {\bm r}}_i & = v_0 {\hat {\bm e}}_i + \textstyle{\sum_{j \neq i}}{\bm u}({\bm r}_{ij}) + {\bf \Gamma}^{-1} \left[-\textstyle{\sum_{j \neq i}}\nabla_{ {\bm r}_i} U({\bm r}_{ij}) \right]; \\
 \dot {\hat {\bm e}}_i & = \left[ \frac{1}{\gamma_r} \textstyle{\sum_{j \neq i}} {\bm M}_{ij} + \textstyle{\sum_{j \neq i}} \bm{\omega}(\bm{r}_{ij}) + \sqrt{2D_{r0}} {\bm \xi}_i \right] \times {\hat {\bm e}}_i.
\end{align}
Here, the anisotropy of the SPR is incorporated through the translational friction tensor $\bm{\Gamma} = \gamma_{\parallel} {\hat {\bm e}}{\hat {\bm e}} + \gamma_{\perp} ({\bf I} - {\hat {\bm e}}{\hat {\bm e}})$, where $\gamma_{\parallel}$ and $\gamma_{\perp}$ represent the translational friction coefficients parallel and perpendicular to the rod's long axis, respectively. The far-field flow ${\bm u}({\bm r}_{ij})$ generated by dipole $j$ advects particle $i$, and the associated angular velocity $\bm{\omega}(\bm{r}_{ij})$ perturbs the orientation of particle $i$. In addition, the steric collisions described by $U({\bm r}_{ij})$ among anisotropic SPRs generate torques ${\bm M}_{ij}$ that further reorient particle $i$. Finally, $\gamma_r$ denotes the rotational friction coefficient, $D_{r0} = 1 / \gamma_r$ is the intrinsic rotational diffusivity of an isolated particle, and ${\bm \xi}_i$ is Gaussian white noise satisfying $\langle {\bm \xi}_i \rangle = {\bm 0}$ and $\langle {\bm \xi}_i(t) {\bm \xi}_j(t') \rangle = {\bf I} \delta_{ij}\delta(t-t')$. The simulations show that $D_{r}$ increases approximately linearly with $\phi$. Single-SPR modeling in the mean-field fashion and many-particle SPR simulations with hydrodynamic interactions switched off are also performed. Further details are provided in the SI Appendix.

\section*{Author contributions}
D.W., M.Y., J.Y., Y.P. Designed the research;
D.W. and Y.L. performed research on strains BW25113 and HCB1284, respectively; L.L. performed modeling;
D.W., R.H., L.L., and Y.L. analyzed data;
D.W., R.H. wrote the paper with input from all authors.

\section*{Competing interests}
The authors declare no competing interest.

\section*{Data availability}
Data used in the study will be available in figshare repository upon publication (DOI: \url{https://doi.org/10.6084/m9.figshare.30907529})

\section*{Acknowledgments}
This work was supported by Fundamental and Interdisciplinary Disciplines Breakthrough Plan of the Ministry of Education of China (JYB2025XDXM502) and National Natural Science Foundation of China Grants (12574238, 12090053, T2325027, 12274448, T2221001, 12474206).

\clearpage

\section*{Supporting Information}

\noindent\textit{Crowding limits relative bacterial surface accumulation with a density-induced effective heating}

\setcounter{figure}{0}
\setcounter{table}{0}
\setcounter{equation}{0}
\renewcommand{\thefigure}{S\arabic{figure}}
\renewcommand{\thetable}{S\arabic{table}}
\renewcommand{\theequation}{S\arabic{equation}}

Here we present the technical details of the methods used to extract the cells' angular diffusivity, both within the $xy$-plane and perpendicular to it. Cross-checks of the results obtained from representative datasets are presented. Additional modeling results supporting the use of a linear dependence between rotational noise and density are also provided. Moreover, we show simulation results based on models beyond the minimal one used in the main text, including another mean-field model using a single self-propelled rod (SPR) and multiparticle simulations---where effective noise emerges from many-body interactions. Finally, a cross-model comparison of the decay of relative surface accumulation is presented.

\section*{Extraction of in-plane rotational diffusivity, $D_{r\!,xy}$}

We characterize the in-plane angular dynamics of bacteria near the bottom wall by calculating the mean square angular displacement (MSAD) of the track direction from entrapped cells. In general, the conventional linear fitting of MSAD works unsatisfactorily. Its applicability is limited to a few experiments and introduces large errors. A demonstration is presented in Fig.~\ref{fig:Drxy_meth}a, left. As mentioned in the main text, to account for measurement uncertainty, circular swimming near the wall, and angular diffusion simultaneously, quadratic polynomial fit is employed [Fig.~\ref{fig:Drxy_meth}(a), center]. Finally, we compute the ensemble average of the single-cell angular autocorrelation function, $\text{ACF}(\delta t) \equiv \langle \bm{n}(\delta t) \cdot \bm{n}(0) \rangle$, where $\bm{n}(t)$ is the unit vector of the cell's instantaneous moving direction. For a rotating motion (circular swimming) superposed with angular white noise, the theoretical ACF decays as $\langle \text{ACF} \rangle = e^{-D_r t} \cos(\Omega t)$~\cite{Zhang2025}. Fitting this functional form to the measured ACF provides a third estimate of $D_{r\!,xy}$ [Fig.~\ref{fig:Drxy_meth}(a), right]. All three methods yield quantitatively consistent results, with systematic variations of ~10\% [Fig.~\ref{fig:Drxy_meth}(b)]. We adopt the quadratic fit for reporting $D_{r\!,xy}$ values in the main text.

The same analysis can be performed using the cell-body angle rather than the trajectory direction; however, the trajectory direction is less sensitive to image-processing noise, and the results from the two approaches are statistically equivalent (Fig.~\ref{fig:Drxy_body}). We therefore use the trajectory angle for all reported $D_{r\!,xy}$ measurements.

\section*{Extraction of polar rotational diffusivity, $D_{r\!,z}$}

In the main text (Methods), we describe two independent methods for extracting the polar rotational diffusivity $D_{r\!,z}$. The first employs a quadratic fit to the MSAD of the polar angle $\theta_z$, analogous to the in-plane analysis. The second models the near-wall angular confinement as a harmonic potential, leading to an analytic expression for the MSAD: $\mathrm{MSAD}(\delta t) = D_{r,z},\tau \big(1 - e^{-2\delta t/\tau}\big)$, where $\tau$ is a characteristic timescale of the confinement. As shown in Fig.~\ref{fig:Drz}, both methods yield equivalent results, with deviations between them of approximately 10\%. This agreement confirms the robustness of the extracted $D_{r\!,z}$ values reported in the main text.

\section*{Applicability of different methods in determining diffusivity}

For the in-plane rotational diffusivity $D_{r,xy}$, quadratic fitting to the mean square angular displacement (MSAD) and exponential fitting to the angular autocorrelation function (ACF) yield equivalent results, while a simple linear fit works poorly due to the existing ballistic contribution from circular motion. For the polar rotational diffusivity $D_{r,z}$, quadratic fitting (using short-to-intermediate times) and the harmonic-potential model (using intermediate-to-long times) also give nearly identical values. In contrast, ACF fitting for $D_{r,z}$ is less reliable because the ACF does not decay to zero---entrapped cells exhibit a non-zero plateau due to angular confinement, see Fig.~\ref{fig:ACF}. To maintain consistency across both measurements, we therefore report values obtained from the quadratic polynomial fit for $D_{r,xy}$ and $D_{r,z}$. The applicability of different methods for extracting $D_{r,xy}$ and $D_{r,z}$ is summarized in Table~\ref{tab:diffusion_methods}.

\begin{table}[htbp]
\caption{Applicability of methods for extracting rotational diffusivities.}
\centering
\begin{tabular}{l c c}
\toprule
\textbf{Diffusivity} & $D_{r,xy}$ & $D_{r,z}$ \\
\midrule
Linear fit to MSAD     & Poorly & --- \\
Quadratic fit to MSAD  & Yes    & Yes \\
Exponential fit to ACF & Yes    & Poorly \\
Harmonic potential fit & ---    & Yes \\
\bottomrule
\end{tabular}
\label{tab:diffusion_methods}
\end{table}

\newpage
\section*{Dependence of rotational noise on density}

We show here that the rotational diffusivity $D_r$ in free space increases nearly linearly with the packing fraction $\phi$, whether the reorientation noise arises from hydrodynamic interactions, from steric collisions, or from both. In the dilute limit [Fig.~\ref{fig:asymptotic}(a)], force dipoles are randomly distributed and induce Gaussian-distributed random torques. The standard deviation $\sigma$ of these random hydrodynamic torques gives $D_r\propto\sigma^2$, showing that $D_r$ increases nearly linearly with $\phi$. When steric collision is the dominant mechanism [Fig.~\ref{fig:asymptotic}(b)], we adopt the self-propelled rod (SPR)~\cite{Bar2020} model and simulate bacteria as hard-sphere dimers. $D_r$ follows $D_r = D_{r0} + \beta\phi/\phi_0$. Lastly, when hydrodynamic interactions and steric collisions are simultaneously present, the linear relation still holds [Fig.~\ref{fig:asymptotic}(c)]. These results, together with the experimentally observed near-linear dependence of $D_{r,xy}$ on $\phi$ [Fig.~2(b) and inset in the main text], support the choice of linear density dependence adopted in our minimal model. Below we provide computational details.

\subsection*{Hydrodynamic-interaction-induced rotational noise}

A swimming cell $j$ is modeled as a force dipole of strength $p$ aligned with its swimming orientation ${\hat {\bm e}}_j$. Such a cell generates a far-field flow $\bm{u}(\bm {r})$ at $\bm{r}$:
\begin{align}\label{eqSM:dipoleFlow}
    {\bm u}({\bm r}) = \frac{p}{8 \pi \eta r^3}\left[ 3 \cos^2 \theta -1 \right] {\bm r},
\end{align}
where $\theta$ denotes the angle between the swimming direction ${\hat {\bm e}}_j$ and the flow position ${\bm r}$. When another cell $i$ is located at $\bm{r}_{ij}$ (its relative position from cell~$j$) with orientation $\hat {\bm e}_i$, this flow ${\bm u}(\bm {r}_{ij})$ reorients cell $i$ with an angular velocity $\bm{\omega}(\bm{r}_{ij})$~\cite{Lauga2009}:
\begin{align}\label{eqSM:dipoleTorque}
    {\bm \omega}(\bm{r}_{ij}) = \frac{1}{2} \nabla \times {\bm u} + \left( \frac{\epsilon^2 - 1}{\epsilon^2 + 1} \right) {\hat {\bm e}}_i \times ({\bm E} \cdot {\hat {\bm e}}_i)
\end{align}
with ${\bm E} = \frac{1}{2}[(\nabla {\bm u}) + (\nabla {\bm u})^{\text T}]$ the rate-of-strain tensor.

To evaluate the enhancement of the effective rotational diffusion coefficient due to hydrodynamic interactions, we consider a target position surrounded by a random distribution of force dipoles of packing fraction $\phi$. The total angular velocity is computed as ${\bm \Omega} = \sum_{j\neq i} {\bm \omega}(\bm{r}_{ij})$. By performing $5 \times 10^7$ independent realizations, we obtain the distribution of ${\bm \Omega}$ for varying $\phi$, as illustrated in Fig.~\ref{fig:asymptotic}(a).

\subsection*{Steric-collision-induced rotational noise}

We simulate a suspension of SPRs with steric collisions (neglecting hydrodynamic interactions) under periodic boundary conditions. The geometric structure and dynamical parameters of SPRs are identical to those detailed in the next section. The motion equations of rod $i$ are given by
\begin{equation}
\label{eqSM:stericMotion}
    {\dot {\bm r}}_i  = v_0 {\hat {\bm e}}_i + {\bf \Gamma}^{-1} \left[-\textstyle{\sum_{j \neq i}}\nabla_{ {\bm r}_i} U({\bm r}_{ij}) \right];
\end{equation}
\begin{equation}
\label{eqSM:stericRotation}
    \dot {\hat {\bm e}}_i  = \left[ \frac{1}{\gamma_r} \textstyle{\sum_{j \neq i}} {\bm M}_{ij} + \sqrt{2D_{r0}} {\bm \xi}_i \right] \times {\hat {\bm e}}_i.
\end{equation}
Here $- \textstyle{\sum_{j \neq i}}\nabla_{ {\bm r}_i} U({\bm r}_{ij})$ represents the total excluded-volume force exerted on rod $i$, and $\textstyle{\sum_{j \neq i}} {\bm M}_{ij}$ is the resulting total torque. Here $D_{r0}$ denotes the constant intrinsic rotational diffusivity of an isolated particle and is defined as $D_{r0} = 1 / \gamma_r$. Moreover, the term ${\bm \xi}_i$ denotes the Gaussian white noise, satisfying $\langle {\bm \xi}_i \rangle = {\bm 0}$ and $\langle {\bm \xi}_i(t) {\bm \xi}_j(t') \rangle = {\bf I} \delta_{ij}\delta(t-t')$. Under varying densities $\phi / \phi_0$, we calculate the autocorrelation function (ACF) of rod orientation, defined as $C(\delta t) = \langle {\hat {\bm e}}(t + \delta t) \cdot {\hat {\bm e}}(t) \rangle_t$, and the corresponding results are shown in Fig.~\ref{fig:asymptotic}(b).

\subsection*{Combined effects on rotational noise}

When hydrodynamic interactions and steric collisions are simultaneously present, we combine the the far-field hydrodynamic coupling and excluded-volume repulsion into the governing equations. For dimer $i$ they read:
\begin{align}
    {\dot {\bm r}}_i & = v_0 {\hat {\bm e}}_i + \textstyle{\sum_{j \neq i}}{\bm u}({\bm r}_{ij}) + {\bf \Gamma}^{-1} \left[-\textstyle{\sum_{j \neq i}}\nabla_{ {\bm r}_i} U({\bm r}_{ij}) \right]; \\
    \dot {\hat {\bm e}}_i & = \left[ \frac{1}{\gamma_r} \textstyle{\sum_{j \neq i}} {\bm M}_{ij} + \textstyle{\sum_{j \neq i}} \bm{\omega}(\bm{r}_{ij}) + \sqrt{2D_{r0}} {\bm \xi}_i \right] \times {\hat {\bm e}}_i.
\end{align}
Here ${\bm u}({\bm r}_{ij})$ and $\bm{\omega}(\bm{r}_{ij})$ are the hydrodynamic translational and rotational velocities, respectively, exerted on rod $i$ by rod $j$, computed via Eqs.~[\ref{eqSM:dipoleFlow}--\ref{eqSM:dipoleTorque}]. All remaining quantities are defined as in Eqs.~[\ref{eqSM:stericMotion}--\ref{eqSM:stericRotation}]. The resulting effective rotational diffusivity $D_r$ is extracted from the ACF and displayed in Fig.~\ref{fig:asymptotic}(c). While the combined rotational noise increases, the linear dependence on $\phi$ remains unchanged.

\clearpage
\section*{Mean-field self-propelled rod model}

Bacteria are frequently modeled as self-propelled rods (SPRs)~\cite{Bar2020} due to their morphological and dynamical similarities. To demonstrate that the minimal model presented in the main text remains qualitatively consistent with a more realistic SPR representation, we examine the steady-state density distribution of a single SPR under an assigned mean-field noise. For simplicity, an SPR is idealized as a rigid dimer consisting of two spheres of diameter $d'$, with a self-propulsion velocity $v_0$ directed along its long axis ${\hat {\bm e}}$, see Fig.~\ref{fig:dimerModel}(a). Following the framework of our minimal model, the dynamics of a SPR at position ${\bm r}$ are described by the following Langevin equations:
\begin{align}
    {\dot {\bm r}} & = v_0 {\hat {\bm e}} + {\bf \Gamma}^{-1} \left[-\nabla_{ {\bm r}} U({\bm r}) \right] + u_z \hat {\bm z}; \\
    -\dot {\alpha} & = \Omega + \frac{1}{\gamma_r} M + \sqrt{2D_{r}} {\xi},
\end{align}
where $-\nabla_{ {\bm r}} U({\bm r})$ denotes the repulsive force exerted by the substrate and $M$ is the resulting torque. $\bm{\Gamma}$ is the anisotropic translational friction tensor, given by ${\bf \Gamma} = \gamma_{\parallel} {\hat {\bm e}}{\hat {\bm e}} + \gamma_{\perp} ({\bf I} - {\hat {\bm e}}{\hat {\bm e}})$. Here $\gamma_{\parallel}$ and $\gamma_{\perp}$ represent the translational friction coefficients parallel and perpendicular to the rod's long axis, respectively. Moreover, $\gamma_r$ is the rotational friction coefficient. In simulation, we set $\gamma_{\parallel} = 4.2 \pi \eta d'$, $\gamma_{\perp} = 4.8 \pi \eta d' $, and $\gamma_r = 5.6 \pi \eta d'^3$~\cite{Lowen1994}. With the exception of the aspect ratio, which is set to $\epsilon = 2$, all other quantities remain identical to those employed in the minimal model.

Using the SPR model, the wall-alignment and the wall-repulsion are naturally enforced through steric collisions between the dimer and the wall, so that no explicit resetting of the polar angle is needed, in contrast to the minimal model. Results of the SPR model are presented in Figs.~\ref{fig:dimerModel}(b) and (c). The accumulation peak in the distribution $\psi(d)$ broadens due to the particle's finite size effect [Fig.~\ref{fig:dimerModel}(b)]. Meanwhile, the weakening trend of relative surface accumulation becomes more gradual, see the dashed line in Fig.~\ref{fig:dimerModel}(c). Note that the decaying trends are qualitatively the same and can be rescaled to match each other, see later in Fig.~\ref{fig:crossmodel}.

\newpage
\section*{Multiparticle simulations}

In the main text, the minimal mean-field model captures the strengthening of cell-cell interactions in denser suspensions through an enhanced effective rotational diffusivity.
Here we go beyond this description with two-dimensional microscopic multiparticle simulations (with fully three-dimensional hydrodynamic interactions), in which cell-to-cell and cell-to-wall interactions are treated explicitly and the intrinsic single-particle diffusivity $D_{r0}$ is held constant. In this case, an increasing effective rotational diffusivity is not set but naturally emerges in denser suspensions, or as more interaction mechanisms are included. The results are again qualitatively the same as those of the minimal model, see Figs.~\ref{fig:MPimage} and \ref{fig:crossmodel}.

However, as more detailed cell-cell interactions are introduced, the simulation results become increasingly sensitive to microscopic parameters (e.g., cell-to-wall distance, higher-order hydrodynamics) that are experimentally unknown. This renders the model less robust and less generic, while being computationally much more demanding. Since it offers no additional physical insight, we retain the minimal model as the primary framework. Computational details of the multiparticle models are given below for completeness.

\subsection*{Hydrodynamic interactions of force dipoles near walls}
We first perform multiparticle simulations accounting only for hydrodynamic interactions. Each swimming cell is modeled as a force dipole subject to Brownian noise, i.e., a spherical ABP with the same settings as the minimal model. $p$ denotes the dipole strength, $v_0$ the self-propulsion speed, and $\hat {\bm e}$ the orientation. The equation of motion for particle $i$ at ${\bm r}_i = [x_i, z_i]$ takes the same form as Eq.~[1] in the main text, supplemented by an additional pairwise hydrodynamic interaction term,
\begin{equation}
\begin{aligned}
    {\dot {\bm r}}_i & = v_0 {\hat {\bm e}}_i + \textstyle{\sum_{j \neq i}}{\bm u}({\bm r}_{ij}) + {\bm u}_i({\bm r}_{i}) -\frac{1}{\gamma_T} \nabla_{{\bm r}_i}U;
\end{aligned}
\label{eqSM:MP_Motion}
\end{equation}
\begin{equation}
\begin{aligned}
    \dot {\hat {\bm e}}_i & = \left[\textstyle{\sum_{j \neq i}} {\bm \omega}({\bm r}_{ij}) + {\bm \omega}_{i}({\bm r}_{i}) + \sqrt{2D_{r0}} {\bm \xi}_i \right] \times {\hat {\bm e}}_i.
\end{aligned}
\label{eqSM:MP_rotation}
\end{equation}
Note that ${\bm u}({\bm r}_{ij})$ and ${\bm \omega}({\bm r}_{ij})$ denote the translational and rotational velocities, respectively, induced on particle $i$ by particle $j$ and the image systems of particle $j$; and ${\bm u}_i({\bm r}_{i})$ and ${\bm \omega}_i({\bm r}_{i})$ are the contributions from image systems of particle $i$ itself.

For a force dipole $j$ at ${\bm r}_j = [x_j, z_j]$ near the bottom wall ($z = 0$), its images at ${\bm r}_j^{\prime} = [x_j, -z_j]$ include an image dipole together with higher-order singularities (e.g., quadrupole)~\cite{Spagnolie2012}. For notational simplicity, in the following equation alone we denote ${\bm r} \equiv {\bm r}_{ij} = {\bm r}_{i} - {\bm r}_{j}$, ${\bm R} \equiv {\bm r}_{i} - {\bm r}_{j}^{\prime}$, ${\bm e} \equiv \hat{{\bm e}}_j$, and $h \equiv z_j$; the translational velocity ${\bm u}({\bm r}_{ij})$ near the bottom wall is then given by
{\small
\begin{equation}
    \label{eqSM:dipoleImages}
    \begin{aligned}
    {\bm u}({\bm r}) &= \frac{p}{8\pi\eta r^3} \left[ 3(\hat{\bm{r}}\cdot\bm{e})^2 - 1 \right] \bm{r} \\
    & +\frac{p}{8\pi\eta R^3} \Big\{ 2e_z R_z \bm{e} - 2e_z(\bm{R}\cdot\bm{e})\hat{\bm{z}} + \left[ \bm{e}\cdot\bm{e}^* - 3(\hat{\bm{R}}^*\cdot\bm{e})(\hat{\bm{R}}\cdot\bm{e}) \right]\bm{R} \Big\} \\
    &+ \frac{3ph^2}{4\pi\eta R^5} \Big[ 2(\bm{R}^*\cdot\bm{e})\bm{e}^* + \bm{R} - 5(\hat{\bm{R}}^*\cdot\bm{e})^2 \bm{R} \Big] + \frac{p h e_z}{2\pi\eta R^3} \Big[ \bm{e}^* - 3(\hat{\bm{R}}^*\cdot\bm{e})\hat{\bm{R}} \Big] \\
    &+ \frac{ph}{4\pi\eta R^3} \Big\{ \left[3(\hat{\bm{R}}^*\cdot\bm{e})^2 - 1\right]\hat{\bm{z}} + 2e_z^* \bm{e}^* - 6(\hat{\bm{R}}^*\cdot\bm{e})\hat{R}_z \bm{e}^* - 6(\hat{\bm{R}}^*\cdot\bm{e})e_z^* \hat{\bm{R}} - 3\hat{R}_z \hat{\bm{R}} + 15(\hat{\bm{R}}^*\cdot\bm{e})^2 \hat{R}_z \hat{\bm{R}} \Big\} \\
    & + \frac{p e_z}{4\pi\eta R^3} \Big[ (\bm{R}^*\cdot\bm{e})\hat{\bm{z}} - R_z \bm{e}^* - e_z^* \bm{R} + 3(\hat{\bm{R}}^*\cdot\bm{e})\hat{R}_z \bm{R} \Big].
    \end{aligned}
\end{equation}
Here the subscript $z$ denotes the $z$-component and the superscript $*$ its sign reversal (e.g., ${\bm R}^* = [R_x, -R_z]$ for ${\bm R} = [R_x, R_z]$), and $\hat{\cdot}$ denotes a unit vector. The corresponding expressions for a dipole near the top wall ($z = H$) follow by the replacement $z \rightarrow H - z$ (note that the signs of the $z$-component of ${\bm e}$ and ${\bm u}$ also inverse). At short range, the flow is regularized by replacing ${\bm r}_{ij}$ with $\ell {\hat {\bm r}}_{ij}$ whenever $r_{ij} < \ell$, where $\ell$ is a cutoff length. The translational velocity reduced by particle $i$'s image on itself, ${\bm u}_i({\bm r}_{i})$, is obtained from Eq.~[\ref{eqSM:dipoleImages}] by setting ${\bm r}_{j} = {\bm r}_{i}$ and omitting the first term on the right-hand side. The rotational velocities ${\bm \omega}({\bm r}_{ij})$ and ${\bm \omega}_{i}({\bm r}_{i})$ are evaluated via Eq.~[\ref{eqSM:dipoleTorque}]. Simulations based on Eqs.~[\ref{eqSM:MP_Motion}--\ref{eqSM:MP_rotation}] yield $\psi_{\text{bottom}} / \psi_{\text{bulk}}$ as a function of density, see Fig.~\ref{fig:MPimage}.

\subsection*{Inclusion of cell-cell scattering}
Furthermore, we incorporate near-field interactions between cells into the model. These interactions are broadly referred to as cell-cell scattering, as they are not limited to steric collisions. We note that rod-shaped bacteria near a wall frequently encounter one another and are thereby reoriented to leave the wall (see Movie~S1). This scattering-induced orientational instability is captured here via an effective alignment interaction (in which head-to-head alignment is unstable). The rotational equation of motion for particle $i$ then takes the same form as Eq.~[\ref{eqSM:MP_rotation}], supplemented by an alignment term:
\begin{align}
    \dot {\hat {\bm e}}_i & = \left[ \textstyle{\sum_{j \neq i}} {\bm \omega}({\bm r}_{ij}) + {\bm \omega}_{i}({\bm r}_{i}) + \textstyle{\sum_{j \neq i}} f(r_{ij})({\hat {\bm e}}_i \times {\hat {\bm e}}_j) + \sqrt{2D_{r0}} {\bm \xi}_i \right] \times {\hat {\bm e}}_i.
\end{align}
Here $f(r_{ij})$ characterizes the strength of the alignment interaction, defined as $f(r_{ij}) = \mathrm{d}U(r_{ij}) / \mathrm{d}r_{ij}$, where $U(r)$ is the Weeks-Chandler-Andersen potential, $U(r)=4\mathcal{E}[\left(\delta/r\right)^{12}-\left(\delta/r\right)^{6}]+\mathcal{E}$ for $r<2^{1/6}\delta$ and $U(r)=0$ otherwise. The characteristic energy and length are $\mathcal{E} = 0.1$ and $\delta = 1$, respectively. To prevent numerical divergence at short range, $f(r_{ij})$ is set to $f(\ell)$ whenever $r_{ij} < \ell$. The equation of translational motion remains identical to Eq.~[\ref{eqSM:MP_Motion}]. Results are displayed in Fig.~\ref{fig:MPimage}. Lastly, it bears emphasis that the cell-cell scattering is much more complex than a single model can capture, owing to the behaviorial variability and many-body interactions, as demonstrated in Movie~S2.

\clearpage
\section*{Cross-model comparison of the accumulation-ratio decay}
\label{sec:crossmodel}
All models above---as well as the three experimental strains---exhibit a decay of $\psi_{\rm bottom}/\psi_{\rm bulk}$ with $\phi$. Here we show that these decays are qualitatively the same, up to a rescaling of the density.

In Fig.~\ref{fig:crossmodel}, taking the wildtype data and the corresponding minimal-model results [Fig.~4(c) in the main text] as the reference, all other curves are shifted horizontally (i.e., with $\phi$ rescaled) to align with them. All data then fall onto a single curve, evidencing a common trend of decay.

The meaning of the rescaling factor depends on the category. For the experiments, it quantifies the per-cell contribution to the collective effective noise: taking the wildtype as reference (factor 1), an elongated cell contributes less (0.75) and a non-tumbling HCB1284 cell the least (0.5), according to the measured $\beta$ ratios.

The same applies to the multiparticle models. The large values (5 for HI; 13 for HI+scattering) reflect that each modeled particle contributes much larger effective noise through the adopted microscopic interactions than an actual wildtype bacterium does. The models create overly strong near-wall interactions because the particle-to-wall distance is much underestimated: the realistic population peaks at 3--5~\textmu{m} above the wall [Fig.~1(b) in the main text; see also~\cite{Wei2025,Htet2024hovering,Burriel2024}], while in the models it peaks below $\sim$2~\textmu{m}. This difference amplifies the near-wall interactions through their higher-order terms in Eq.~[\ref{eqSM:dipoleImages}].

In the mean-field models, the interpretation differs, because $\phi$ enters only through the assigned $D_r(\phi)$ rather than through explicit interactions. The scaling factor of $\phi$ therefore measures how strongly the modeled mechanism at a given noise level can entrap a particle. The SPR is more retentive than the point particle---owing to its elongated geometry and the absence of polar-angle resetting---such that its accumulation at density $\phi$ matches that of the minimal model at $\phi/3$ (factor 0.33).

The collapse in Fig.~\ref{fig:crossmodel} thus summarizes the overall message: whatever the microscopic mechanism, many-body interactions under crowding can be equivalently converted into an effective noise level that dictates the decay of relative surface accumulation. Finally, it is worth noting that the detailed models contain multiple experimentally inaccessible parameters (e.g., the particle-to-wall distance of individual cells, the cell-cell scattering strength). Their inclusion adds little new insight into the weakening of relative surface accumulation studied here, and the sensitivity of the results to these parameters undermines the robustness of the modeling. Such microscopic details are moreover system-specific---collision rules, hydrodynamic coupling, and cell geometry all vary across microswimmers---limiting the general applicability of the detailed models. We therefore leave out these details and retain the minimal model as the primary framework.
\clearpage
\section*{Movie Captions}
\begin{description}
\item[Movie S1.] \textbf{Cell-cell scattering helps an entrapped cell escape.} Top view of typical scattering-induced escape events. To aid visualization, each panel is paused briefly at the moment of cell crossing, with the crossing site marked by a red circle.
\item[Movie S2.] \textbf{Complex cell-cell interactions within the entrapped layer.} Side view of cell-cell interactions near the surface, demonstrating the variety of interaction outcomes. The dashed line marks the surface in each panel. (1) Two cells collide head-on, and one reverses afterward. (2) Two cells swim side by side, and lateral interactions help one cell leave. (3)--(4) An increasing number of cells interact simultaneously. (5) A cell approaching the surface reverses and leaves directly, without being entrapped.
\end{description}

\begin{figure}
\centering
\includegraphics[width=1\textwidth]{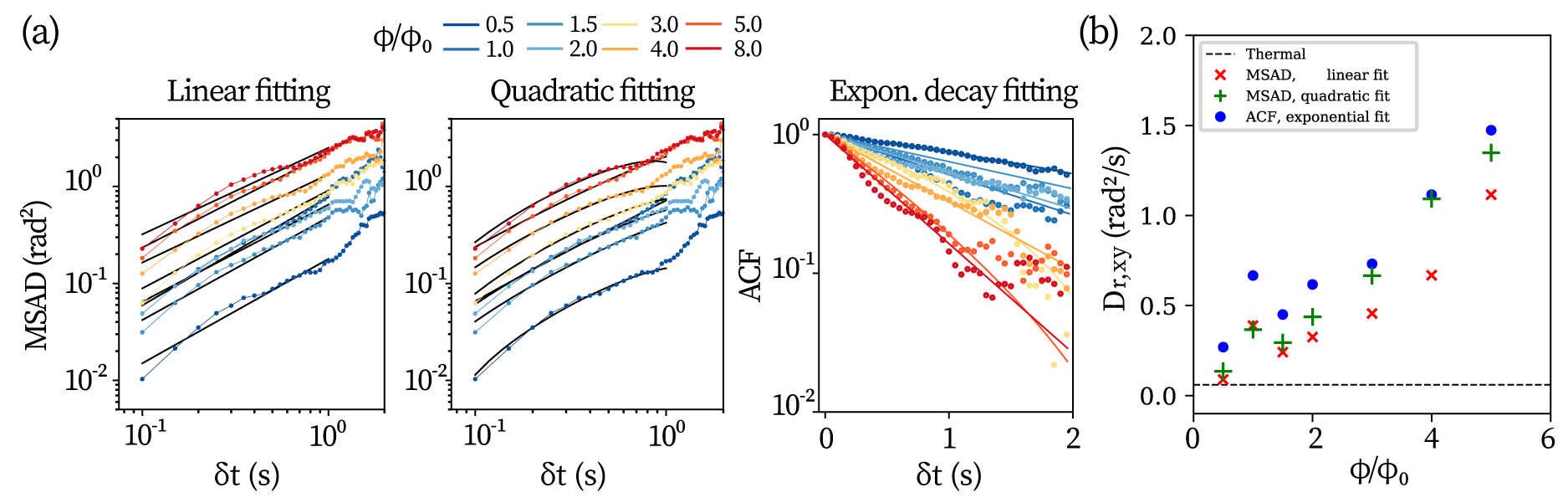}
\caption{\textbf{Comparison of in-plane rotational diffusivity extraction methods.} (a) Extracting $D_{r\!,xy}$ via a linear fit and a quadratic fit to the mean square angular displacement (MSAD), or by fitting an exponential decay to the angular autocorrelation function (ACF). A representative dataset is shown. (b) Comparison of $D_{r\!,xy}$ values obtained from the three methods.}
\label{fig:Drxy_meth}
\end{figure}

\begin{figure}
\centering
\includegraphics[width=0.75\textwidth]{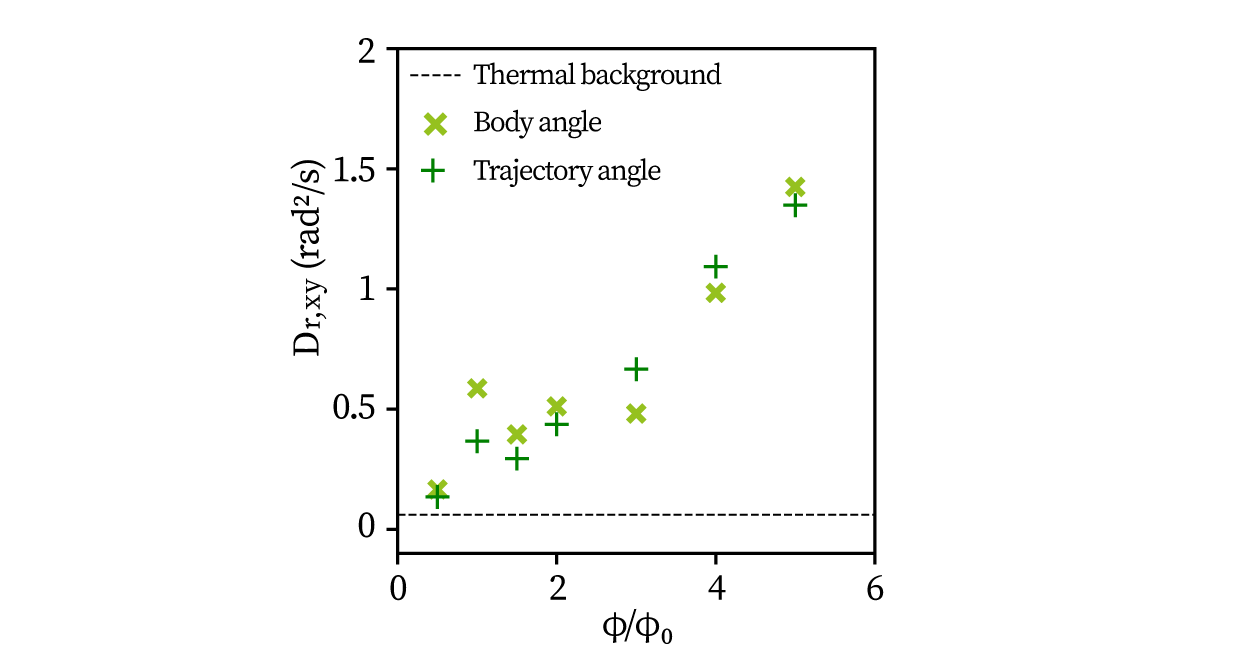}
 \caption{\textbf{Comparison of results using cell-body angle vs. trajectory angle.} The $D_{r\!,xy}$ values obtained via quadratic fitting are compared for the same dataset shown in Fig.~\ref{fig:Drxy_meth}.}
\label{fig:Drxy_body}
\end{figure}

\begin{figure}
\centering
\includegraphics[width=0.75\textwidth]{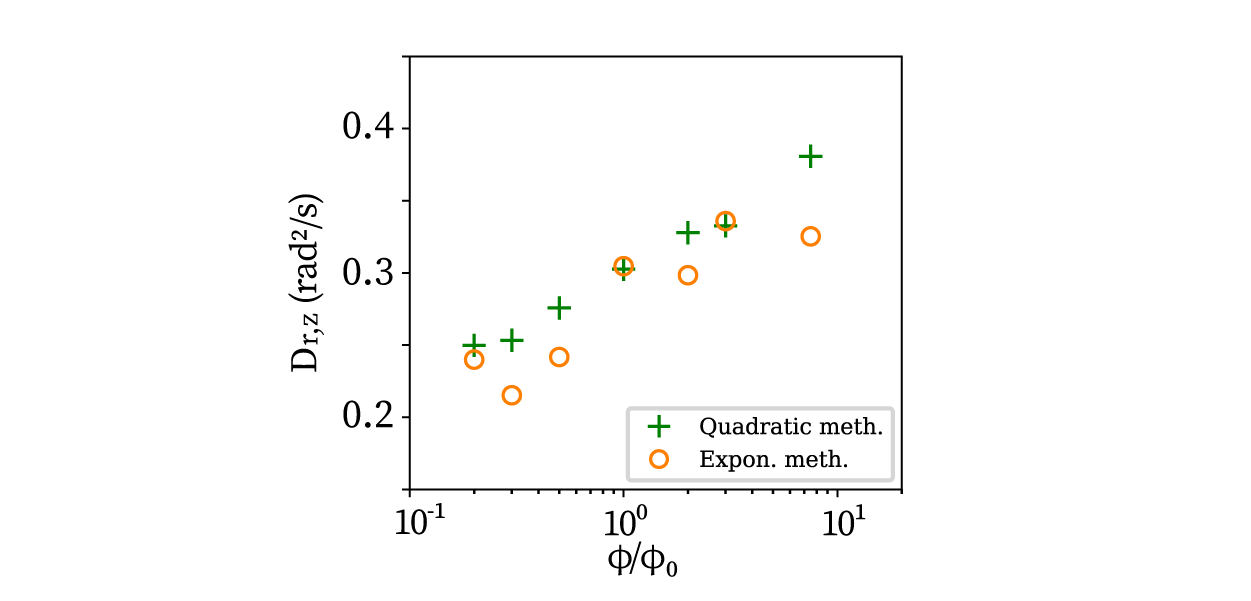}
\caption{\textbf{Comparison of polar rotational diffusivity extraction methods.} $D_{r\!,z}$ values obtained from a quadratic fit to the MSAD are compared with those from an exponential fit to $\mathrm{MSAD}(\delta t) = D_{r\!,z}\,\tau \big(1 - e^{-2\delta t/\tau}\big)$.}
 \label{fig:Drz}
\end{figure}

\begin{figure}
\centering
\includegraphics[width=0.75\textwidth]{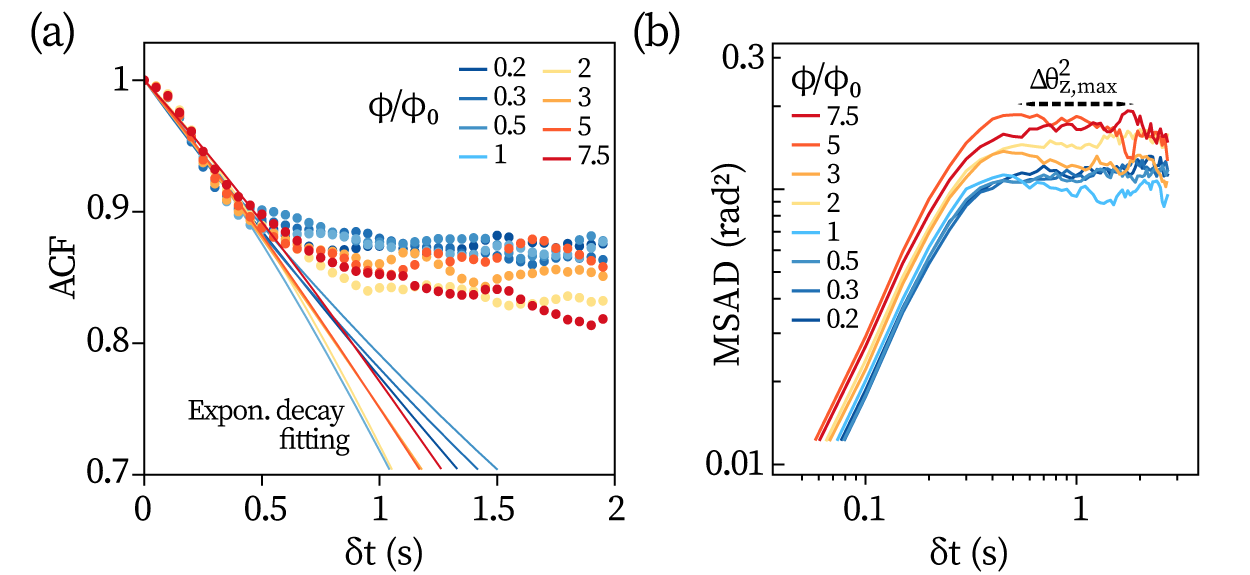}
\caption{\textbf{ACF of $\theta_z$ for entrapped cells.} (a) Angular autocorrelation functions (ACFs) of entrapped elongated cells at varying densities. Lines show exponential fits to the initial decay. (b) MSAD of the same dataset, shown in the main text as Fig.~2(c).}
 \label{fig:ACF}
\end{figure}

\begin{figure}
\centering
\includegraphics[width=0.7\textwidth]{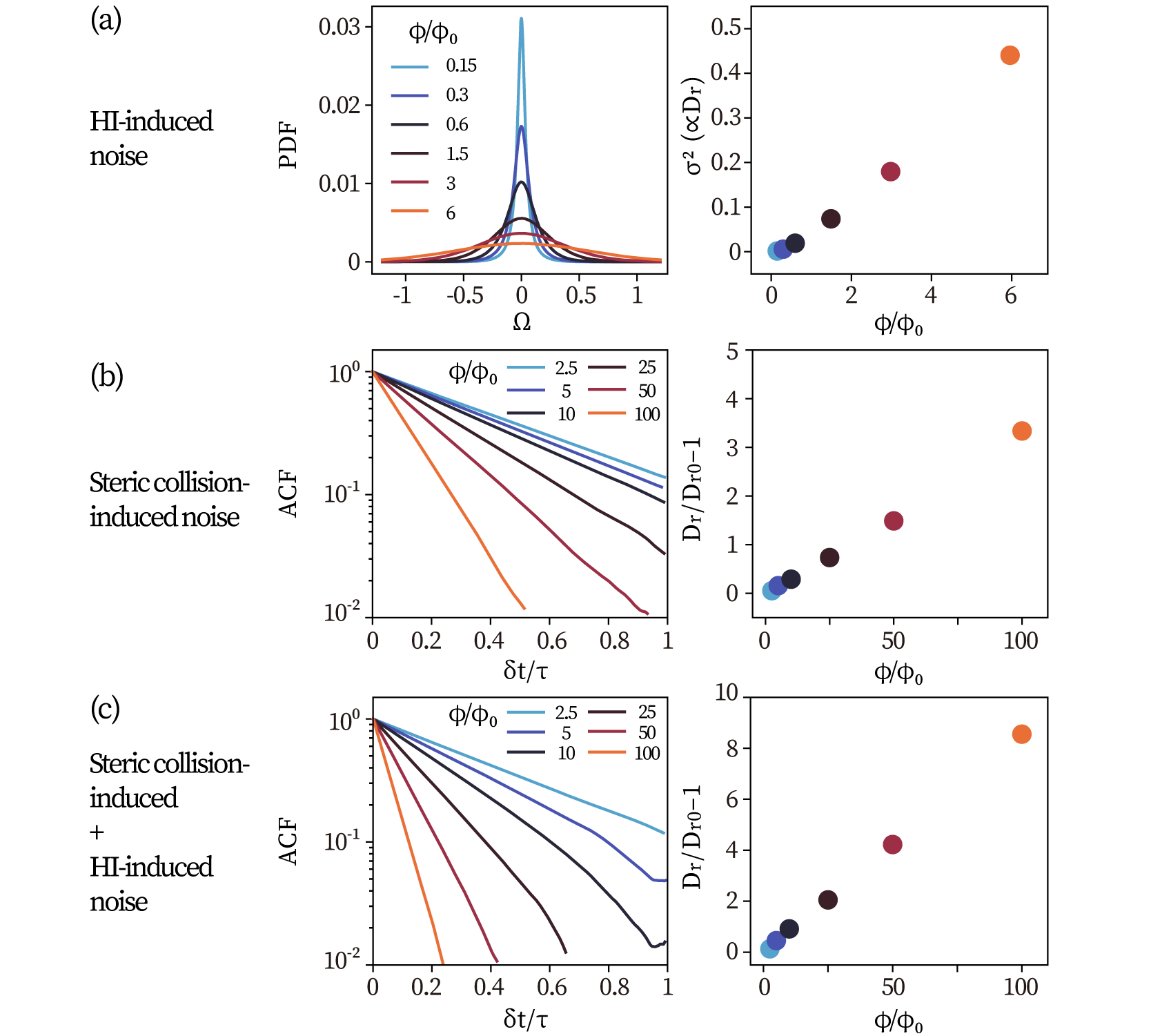}
\caption{\textbf{Free-space rotational diffusivity $D_r$ versus density.} (a) Dilute regime, where reorientation arises from hydrodynamic interactions. The scaled packing fraction $\phi/\phi_0 = n V_{\rm cb} / \phi_0$, with $n$ the number density of force dipoles, $V_{\text{cb}}$ the cell-body volume, and $\phi_0 = 0.19\%$ the unit packing fraction defined in the main text. The standard deviation $\sigma$ of the random reorientations is obtained from Gaussian fits. (b) Dense regime with steric collisions only, simulated using hard-sphere dimers (multiparticle SPR model with rod-rod collisions, see text). $D_{r0}$ denotes the isolated single-particle rotational diffusivity, and $\tau = 1/D_{r0}$. (c) Same as (b), but with hydrodynamic interactions additionally included.}
 \label{fig:asymptotic}
\end{figure}

\begin{figure}
\centering
\includegraphics[width=0.95\textwidth]{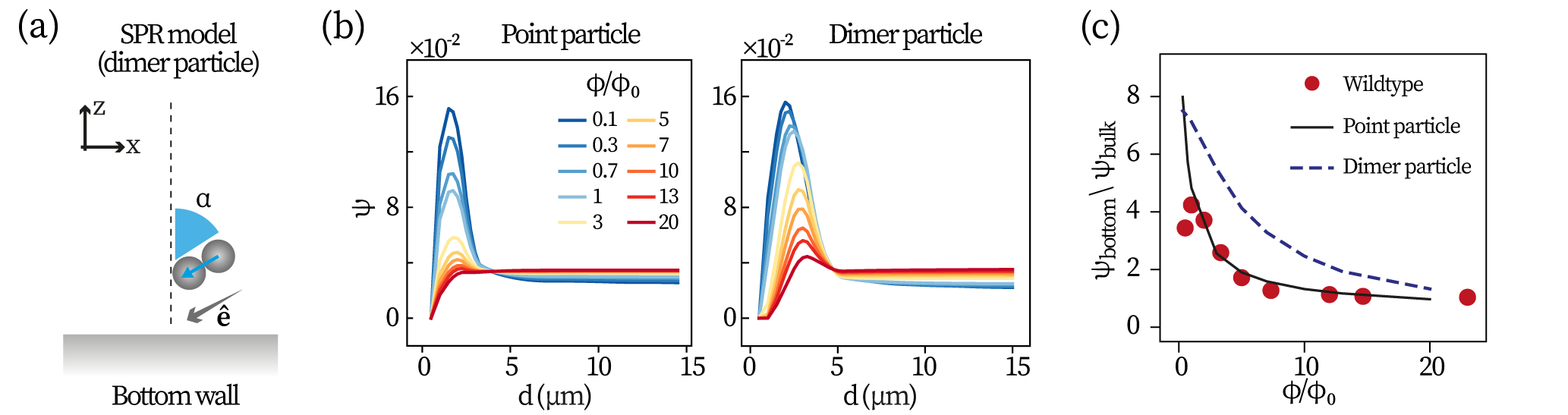}
\caption{\textbf{Results of the additional self-propelled rods (SPR) model.} (a) Schematic of the model: a dimer of hard spheres represents the SPR (i.e., a rod-shaped bacterium). Anisotropy in the translational drag coefficient, wall alignment, and wall repulsion are naturally incorporated through particle--wall collisions in this scheme. Other conditions are kept the same as in the minimal model. (b) Normalized population distribution $\psi(d)$ under varying concentrations. Left: minimal model results [from Fig.~4(b) in the main text]; right: SPR model results. (c) Relative surface accumulation $\psi_{\rm bottom}/\psi_{\rm bulk}$ from the SPR model compared with the results presented in the main text.}
 \label{fig:dimerModel}
\end{figure}

\begin{figure}[p]
\centering
\includegraphics[width=0.6\textwidth]{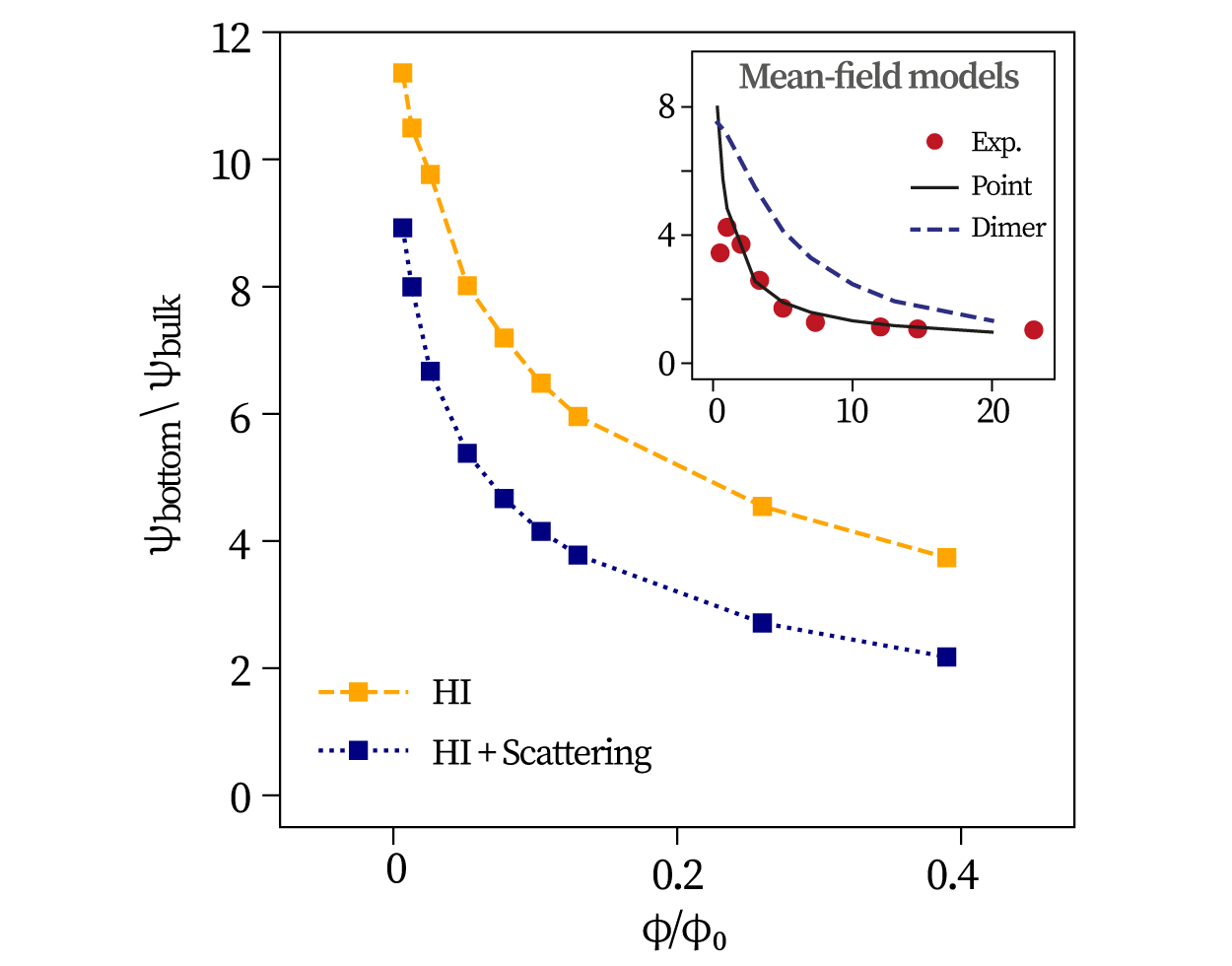}
\caption{\textbf{Surface accumulation in near-wall multiparticle simulations.} Relative accumulation ratio $\psi_{\rm bottom}/\psi_{\rm bulk}$ as a function of density $\phi$, for simulations accounting only for hydrodynamic interactions (HI) and hydrodynamic interactions supplemented with cell-cell scattering (HI+scattering). Both reproduce the decay of the accumulation ratio with increasing density. While the decay proceeds faster than in the mean-field models and experiments [inset, same as Fig.~\ref{fig:dimerModel}(c), shown for visual comparison], the trends are similar (see the cross-model comparison in Fig.~\ref{fig:crossmodel}). The faster decay is attributed to the systematically underestimated particle-to-wall distance (see text).}
\label{fig:MPimage}
\end{figure}

\begin{figure}[p]
\centering
\includegraphics[width=0.75\textwidth]{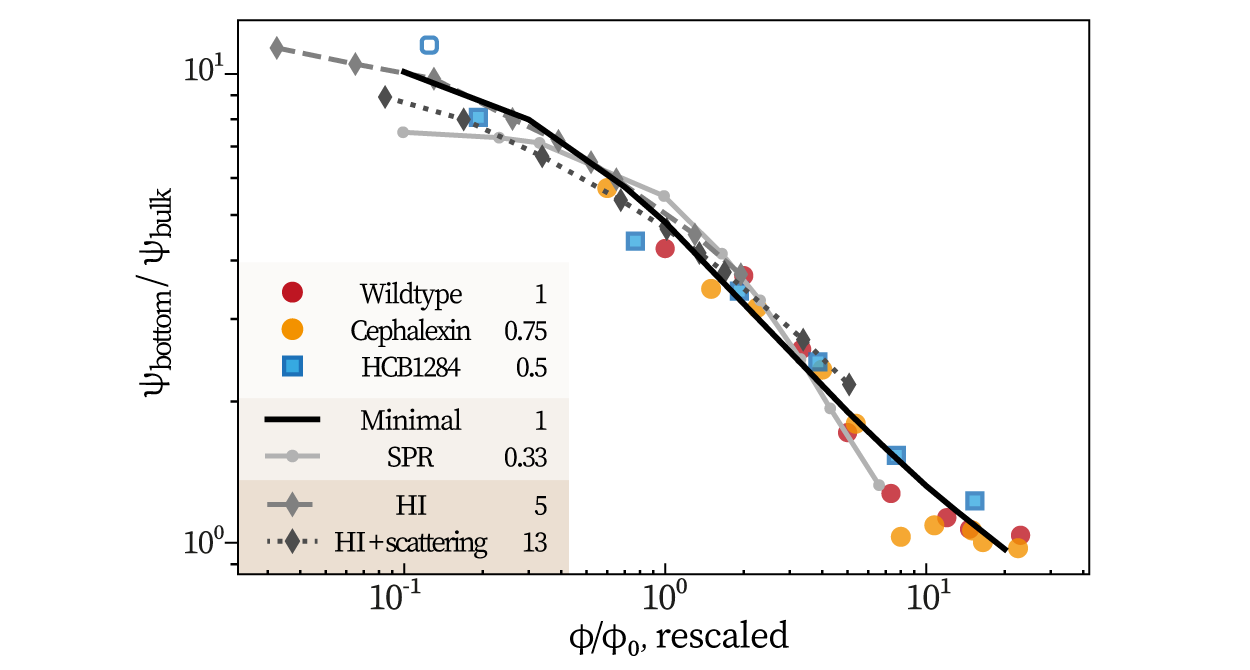}
\caption{\textbf{Comparison of the accumulation-ratio decay across strains and models.} $\psi_{\rm bottom}/\psi_{\rm bulk}$ versus density for the three experimental strains, the mean-field models with an assigned $D_r(\phi)$ (Minimal; SPR), and the multiparticle models with emergent noise (HI; HI+scattering), each curve shifted only along the horizontal axis (density $\phi$) by the factor indicated in the legend. Factors for experiments are obtained from the measured $\beta$ ratios [1:0.75:0.5, see Fig.~4(c) in the main text]; model factors reflect the per-particle contribution to the effective noise strength (see text). The lowest-density point of each tumbling strain, which does not belong to the density-induced decay, is omitted.}
\label{fig:crossmodel}
\end{figure}

\FloatBarrier

\bibliography{reference}

\end{document}